\documentclass[a4paper,superscriptaddress,english,aps,twocolumn,floatfix, showpacs, amsfonts, amssymb,prb]{revtex4-1}
\usepackage{epsfig,psfrag,amsmath,amssymb,float}
\usepackage[T1]{fontenc}
\usepackage[latin9]{inputenc}
\usepackage{amsmath}
\usepackage{amssymb}
\usepackage{color}
\usepackage{amscd}
\usepackage{bm}
\usepackage{psfrag}

\usepackage[bookmarks=true,colorlinks,linkcolor=blue,urlcolor=blue,citecolor=blue]{hyperref}
\usepackage{babel}

\newcommand{\be}{\begin{equation}}
\newcommand{\ee}{\end{equation}}
\newcommand{\bea}{\begin{eqnarray}}
\newcommand{\eea}{\end{eqnarray}}

\newcommand{\mc}{\mathcal}

\def\nn{\nonumber\\}
\def\fr#1{(\ref{#1})}
\def\up{\uparrow}
\def\down{\downarrow}
\def\eps{\epsilon}

\begin{document}
\title{Spin-charge separated quasiparticles in one dimensional quantum fluids}
\author{F. H. L. Essler}
\affiliation{The Rudolf Peierls Centre for Theoretical Physics, Oxford
University, Oxford OX1 3NP, UK} 
\author{R. G. Pereira}
\affiliation{Instituto de F\'{i}sica de S\~ao Carlos, Universidade de
S\~ao Paulo, C.P. 369, S\~ao Carlos, SP,  13560-970, Brazil} 
\author{I. Schneider}
\affiliation{Department of Physics and Research Center OPTIMAS,
University of Kaiserslautern, D-67663 Kaiserslautern, Germany}
\date {\today}
\begin{abstract}
We revisit the problem of dynamical response in spin-charge separated
one dimensional quantum fluids. In the framework of Luttinger liquid
theory, the dynamical response is formulated in terms of
noninteracting bosonic collective excitations carrying either charge
or spin. We argue that, as a result of spectral nonlinearity,
long-lived excitations are best understood in terms of generally
strongly interacting fermionic holons and spinons. This has far
reaching ramifications for the construction of mobile impurity models
used to determine threshold singularities in response functions. We
formulate and solve the appropriate mobile impurity model describing
the spinon threshold in the single-particle Green's function. Our
formulation further raises the question whether it is possible to
realize a model of noninteracting fermionic holons and spinons in
microscopic lattice models of interacting spinful fermions. We
investigate this issue in some detail by means of density matrix
renormalization group (DMRG) computations.

\end{abstract}

\pacs{71.10.Pm,71.10.Fd}

\maketitle

\section{Introduction}
Understanding the essential features of
a quantum  many-body system usually entails finding a simple
explanation of its low-energy spectrum in terms of weakly interacting,
long-lived quasiparticles. For example, in Landau's Fermi liquid
theory the quasiparticles are fermions carrying the same quantum
numbers as an electron, namely charge $e$ and spin
1/2. However, it is  well established that the long lived excitations
in strongly correlated systems may carry only a fraction of the
quantum numbers of the elementary constituents. In fact,
fractionalization is  often invoked as a route towards exotic phases
of matter such as spin liquids or high-temperature superconductors \cite{xu}. 

Perhaps the most prominent example of fractionalization is spin-charge
separation in one dimensional (1D) quantum fluids known as Luttinger liquids
\cite{deshpande}. The hallmark of these theories is a low energy
spectrum described by two decoupled free bosonic fields associated with
collective spin and charge degrees of freedom, respectively.
On the experimental side, the most direct evidence for spin-charge
separation involves the observation of multiple peaks associated with
spin and charge collective modes in dynamical response functions at
fairly high energies \cite{auslaender,kim,jompol,schlappa}, beyond the
regime where Luttinger liquid theory is applicable.
Spin-charge separation is known to persist at high energies
for integrable theories such as the 1D Hubbard 
\cite{hubbardbook,EK94a,EK94b,Andrei,Deguchi,penc1,favand,penc,benthien,adrian} and $1/r^2$
$t$-$J$\cite{arikawa01,arikawa04} models. In these cases
any eigenstate with a finite energy in the thermodynamic limit can be
classified in terms of elementary excitations called holons (which
carry charge $e$ and spin 0) and spinons (which carry charge 0 and spin
$1/2$), along with their bound states. A convenient way for describing
such excitations as well as their scattering is to define corresponding
creation and annihilation operators $Z^\dagger_a(\theta)$ and $Z_a(\theta)$,
where $a$ labels the different types of elementary excitations and
$\theta$ is a variable that parameterizes the momentum
$p_a(\theta)$ (the form of this function depends on the particular
model under consideration). As a consequence of integrability, creation
and annihilation operators fulfil the Faddeev-Zamolodchikov algebra
\cite{Zam,Fadd} 
\bea
Z_{a}(\theta_1)Z_{b}(\theta_2)&=&
S_{ab}^{cd}(\theta_1,\theta_2)Z_{d}(\theta_2)Z_{c}(\theta_1),\nn
Z_{a}(\theta_1)Z^\dagger_{b}(\theta_2)&=&2\pi\delta(\theta_1-\theta_2)
\delta_{a,b}\nn
&&+
S_{bc}^{da}(\theta_2,\theta_1)Z^\dagger_{d}(\theta_2)Z_{c}(\theta_1).
\eea
Here $S_{ab}^{cd}(\theta_1,\theta_2)$ is the purely elastic two-particle
S-matrix. The corresponding elementary excitations are infinitely
long lived, but generally strongly interacting, as can be seen from
their scattering matrices \cite{EK94a,EK94b,Andrei}. Moreover, their
quantum numbers, e.g. charge 0 and spin 1/2 for spinons in the Hubbard
model, differ from those of the collective bosonic spin modes in the
Luttinger liquid description. It is natural to assume that breaking
integrability slightly will not generically lead to a qualitative
change in the nature of elementary holon and spinon excitations, but
merely render the lifetimes finite. 

How to reconcile this picture emerging from the exact solution with
Luttinger liquid theory? In the latter, the nature of elementary
excitations is obscured by the fact that, due to the linear dispersion
approximation, the spectrum is highly degenerate and allows for many
interpretations, which ultimately all give the same results for
physical observables. Examples are chiral Luttinger liquid
descriptions of quantum Hall edges \cite{kareljan} and the low-energy
excitations of the Hubbard model. The latter can be understood both in
terms of interacting, fermionic holons and spinons
\cite{JNW,hubbardbook,woynarovich,frahmkorepin}, and in terms of
noninteracting bosons associated with collective spin and charge
degrees of freedom \cite{Affleck,gogolin,giamarchi}. 

Going beyond the linear dispersion approximation is expected to remove
ambiguities in the quasiparticle description of the spectrum: there
will be a particular choice that maximizes the lifetimes of
elementary excitations. In integrable models such as the Hubbard chain
these lifetimes are infinite.

Over the last decade it has been established in a series of works that
going beyond the linear dispersion approximation is essential for
correctly describing the dynamics of one dimensional models with gapless excitations
\cite{work1,rozhkov,Roz06,work2,carmelo1,carmelo2,carmelo3,BAW1,work3,work4,work5,IG08,work6,BAW2,ABW,work7,work8,PWA09,pereira12,Roz14,austen,PS,Pereira,SIG1,SIG2,FHLE,SIG3,ts,Seabra}. This
has resulted in the so-called ``nonlinear Luttinger liquid'' (nLL) approach to
gapless 1D quantum liquids; see Refs.~\onlinecite{SIG3,pereira12} for recent
reviews. The basic reason for the failure of linear Luttinger liquid
theory is that at finite energies the running coupling constants of
irrelevant perturbations such as band curvature terms are in fact
different from zero. Taking them into account in perturbation theory
leads to infrared singularities. These need to be resummed to all
orders in the coupling constants, which gives rise to new,
momentum-dependent exponents in response functions.  

A key ingredient of the nLL method is the identification of quasiparticles
describing excited states both at high and at low energies. This is
straightforward for spinless fermions, because the quasiparticles are
adiabatically connected to free fermions \cite{PWA09}. The spinful
case is considerably more involved due to 
the onset of spin-charge separation for arbitrarily weak interactions
and the concomitant qualitative change in the nature of the elementary
excitations compared to the noninteracting limit. It was realized in
Refs.~\onlinecite{PS,SIG1,SIG2,Pereira} that in many important cases
exact results can be obtained by using a phenomenological model of
weakly interacting fermionic holons and spinons, whose dispersions
delineate the edges of the support of the dynamical response function
under consideration. By construction these excitations are different from
the true elementary excitations in the Hubbard model. In particular,
they carry different spin and charge quantum numbers. It is then an
obvious but crucial question how to reconcile this approach with
the exact solution of integrable models like the Hubbard chain.

In this work we develop a new approach to deriving mobile impurity
models for studying dynamical correlations in gapless models of
spinful fermions. We first carry out a direct construction of the
``physical'' holon and spinon fields in the limit of weak electron-electron interactions. This results in a representation of spinful
nonlinear Luttinger liquids in terms of strongly interacting holons
and spinons, which is in direct accord with known results obtained
in integrable cases. Recalling that the spinless fermion case was best
understood by considering weak interactions, we address the
construction of a microscopic model of interacting electrons giving
rise to a theory of \emph{noninteracting} fermionic holons and
spinons at low energies, but in presence of spectral
nonlinearities. We then   derive mobile impurity models to
analyze dynamical response functions in our formulation. We
demonstrate explicitly how to recover results obtained previously by
means of the approach of Refs.~\onlinecite{SIG1,SIG2}.
Finally, we address the question how to realize a lattice model of
interacting electrons that gives rise to noninteracting holons and
spinons at low energies.

\section{Spinless Fermions}
\label{sec:spinless}

Before we approach  the problem of defining  quasiparticles for spin-$1/2$ fermions, it is instructive to review the case of spinless fermions. For concreteness, consider the simplest lattice model of  interacting spinless fermions: \be
H=-t\sum_{j=1}^L(c^\dagger_jc^{\phantom\dagger}_{j+1}+\textrm{h.c.})+V\sum_{j}n_jn_{j+1}.\label{tVmodel}
\ee
Here $c_j$ is the annihilation operator for a fermion at site $j$,
$n_j=c^\dagger_jc^{\phantom\dagger}_j$ is the number operator, $t$ is
the hopping parameter, and $V$ is the  nearest-neighbor interaction
strength. This model has a U(1) symmetry, $c_j\to e^{i\alpha}c_j$, $\alpha\in \mathbb R$, 
associated with conservation of the total number of fermions $N$.
Moreover, the model is integrable and the exact spectrum for arbitrary
values of $V$  can be calculated  from the Bethe ansatz solution
\cite{korepinbook,Faddeev84}. There is  a gapless phase extending between
$-2<V\leq 2$. At  $V=2$ the model is equivalent to the SU(2)-symmetric
Heisenberg spin chain \cite{korepinbook}. 

A useful starting point for analytical approximations is the
noninteracting  model with $V=0$. In this case, the elementary
excitations are free fermions with dispersion relation $\epsilon_0(k)=
-2t\cos k-\mu $ measured from the Fermi energy $\mu$. Low-energy
excitations are particles and holes with momentum close to the Fermi
points, $k\approx \pm k_F$. The Fermi momentum is related to the
average density by $k_F=\pi N/(La_0)$, where $a_0$ is the lattice spacing.  

At weak coupling, $V/t\ll1$, standard  bosonization can be used to
derive an effective low-energy theory for model \eqref{tVmodel}, see
e.g. Refs. \onlinecite{Affleck,gogolin,review}. One
starts by taking the continuum limit and projecting the fermionic
field onto states with   momentum   near the Fermi points. This leads
to the right- and left-moving components in the mode expansion
\be 
c_j\to \sqrt{a_0} [e^{ik_F x}R(x)+e^{-ik_F x}L(x)].\label{modeexpansion}
\ee 
The effective Hamiltonian in terms of $R$ and $L$ fermions reads
\bea
\mc H&=&\int dx [-  v^\prime R^\dagger i\partial_xR+  v^\prime
  L^\dagger i\partial_xL\nonumber\\ 
&&+gR^\dagger RL^\dagger L+\mc H_{\textrm{irr}}(x)],\label{Hfermion}
\eea
where all operators are normal ordered with respect to the
noninteracting Dirac sea. To first  order in $V$, we have the
parameters  $ v^\prime\approx 2ta_0\sin k_F\left(1+\frac{V}{\pi t}\sin
k_F\right)$ and ${g\approx 4Va_0\sin^2 k_F}$. The coupling constant
$g$ is the only  marginal interaction. The term $\mc H_{\textrm{irr}}(x)$ contains 
nonlinear dispersion terms and other interactions which are irrelevant
in the renormalization group (RG) sense. 
In the standard Luttinger
liquid approach this term is dropped, which corresponds to linearizing
the dispersion around the Fermi points $\pm k_F$.

The bosonization formula for chiral spinless fermions reads\bea
R(x)&=&\frac{\eta}{\sqrt{2\pi }} \, e^{-\frac{i}{\sqrt2}\varphi(x)},\\
L(x)&=&\frac{\bar \eta}{\sqrt{2\pi}} \, e^{\frac{i}{\sqrt2}\bar\varphi(x)},
\eea
where $\eta,\bar \eta$ are Majorana fermions and $\varphi(x),\bar \varphi(x)$ are chiral bosons that obey the commutation relations\bea
[\varphi(x),\bar \varphi(y)]&=&0,\\
{[\varphi(x),\varphi(y)]}&=&2\pi i\text{sgn}(x-y)=-[\bar\varphi(x), \bar\varphi(y)].
\eea
Throughout this paper we use ``CFT normalizations'' for bosonic vertex operators
\be
\langle e^{i\alpha\varphi(x)}e^{-i\alpha\varphi(y)}\rangle=
\frac{1}{(x-y)^{2\alpha^2}}.
\label{normalization}
\ee
A consequence of employing these conventions is that vertex
operators are dimensionful
\be
{\rm dim}\left(e^{i\alpha\varphi(x)}\right)={\rm length}^{-\alpha^2}.
\ee
We also define the canonical  bosonic field $\Phi(x)$ and its dual
$\Theta(x)$ by 
\bea
\Phi(x)&=&\varphi(x)+\bar \varphi(x),\\
\Theta(x)&=&\varphi(x)-\bar \varphi(x),
\eea
which obey  \be
[\Phi(x),\Theta(x^\prime)]=4\pi i\text{sgn}(x-x^\prime).\ee
Bosonization of Eq. \eqref{Hfermion} then leads to the Hamiltonian  \be
\mc H=\int dx [\mc H_{\textrm{LL}}(x)+\mc H_{\textrm{irr}}(x)].\label{HLLHIrr}
\ee
The first term,
\be
\mc H_{\textrm{LL}}(x)=\frac{v}{16\pi} \left[K(\partial_x\Theta)^2+\frac1K(\partial_x\Phi)^2\right],\label{HLL}
\ee
is  the Luttinger model. The velocity $v$ and the Luttinger parameter
$K$ are given to first  order in $V$ by $v\approx  v^\prime$ and
$K\approx 1-\frac{V}{\pi t}\sin k_F$.  
The bosonic fields describe the collective low-energy density mode of the
quantum fluid. The uniform part of the 
density operator is related to $\Phi(x)$ by the bosonization relation
\be 
Q(x)=R^\dagger (x)R(x)+L^\dagger(x)L(x)\sim - \frac{1}{\pi\sqrt8}\partial_x\Phi(x).
\ee
The total  charge operator is given by the integral \be
q=\int_{-\infty}^{+\infty} dx\,Q(x).\label{charge}
\ee
Thus, an elementary excitation with charge $q=1$ corresponds to a kink
of amplitude $\pi\sqrt8$ in the bosonic field $\Phi(x)$. As is well
known,  the  model in Eq. (\ref{HLL}) correctly captures the
long-distance asymptotic decay of correlation functions for any
1D system in the Luttinger liquid universality class
\cite{Affleck,gogolin,giamarchi}.

The limitations of Luttinger liquid theory appear when one considers
dynamical response functions at small but finite frequency and
momentum. At finite energy scales it becomes necessary to take the
irrelevant perturbations ${\cal H}_{\rm irr}$ to the Luttinger model
into account. At weak coupling and in the
absence of particle-hole symmetry (i.e. away from half-filling in the
lattice  model), the leading corrections in  (\ref{Hfermion}) are the
dimension-three operators
\bea
\mc H_{\textrm{irr}}&=&-\frac1{2\tilde m}(R^\dagger\partial_x^2R +L^\dagger\partial_x^2L) \nonumber\\
&&+g_3 (R^\dagger RL^\dagger i \partial_x L-L^\dagger LR^\dagger
i\partial_x R+\textrm{h.c.}).
\label{irretas}
\eea
Here we have introduced $\tilde m^{-1}\approx
m^{-1}+\frac{Va_0^2}{\pi}\sin 2k_F$, with $m^{-1}=2ta_0^2 \cos k_F$
the inverse free fermion mass, and $g_3\approx Va_0^2\sin
2k_F$. Bosonizing these irrelevant  terms we obtain cubic
boson-boson interactions
\bea 
\mc H_{\textrm{irr}}&=&\lambda_3^+ \partial_x\Phi[(\partial_x\Phi)^2+(\partial_x\Theta)^2]\nonumber\\
&&+\lambda_3^- \partial_x\Phi[(\partial_x\Phi)^2-(\partial_x\Theta)^2],
\label{phicubed}
\eea
with $\lambda_3^+\approx
-\frac{1}{48\sqrt{2}\pi}\left[\frac1m+\frac{g_3}{\pi}\right]$ and
$\lambda_3^-\approx
\frac{1}{96\sqrt2\pi}\left[\frac1m-\frac{2g_3}{\pi}\right]$. 
 While the bosonic representation allows one to take the marginal
interaction $g$ into account exactly, perturbation theory
in the nonlinear boson interactions \fr{phicubed} suffers from
infrared divergences \cite{work2,pereira12}.  The latter are
associated with the huge degeneracy of states in the linear dispersion
approximation: all states with an arbitrary number of bosons moving in
the same direction carrying the same total momentum  are degenerate
\cite{work2,pereira12}. 
 
A way to circumvent these difficulties in analyzing the
nonlinear bosonic theory is suggested by reverting to the fermionic
representation (\ref{irretas}).  At the free fermion point, the
irrelevant interaction vanishes, $g_3=0$, and one is left with the
quadratic dispersion term with effective mass $m$. 
Taking the nonlinear dispersion into account in the free fermion model
removes the degeneracy of particle-hole pairs that carry the same total
momentum. One can then approach the problem from free fermions with
nonlinear dispersion and include interactions perturbatively
\cite{work1}. This approach reveals that the most pronounced effect of
the interactions is to give rise to power-law singularities at the
edges of the excitation spectrum. While a complete analytical
solution of model (\ref{Hfermion}) taking into account both band
curvature and interaction is highly nontrivial, the edge singularities
can be described by an effective impurity model in analogy with the
x-ray edge problem \cite{work1,Schotte}.  

Consider, for instance,    the single-fermion spectral function
\be 
A(\omega,k)=-\frac{1}{\pi}\text{Im }G_{\text{ret}}(\omega,k),
\ee
where\bea
G_{\text{ret}}(\omega,k)&=&-i\int_0^{\infty}dt\, e^{i\omega t}\sum_l\, e^{-i k l a_0}\nonumber\\
&&\times\langle \psi_0|\{c^{\phantom\dagger}_{j+l}(t),c^{ \dagger}_{j}\}|\psi_0\rangle\label{Greensfunction}
\eea
is the retarded  Green's function, with $|\psi_0\rangle$ the exact ground state. We can separate the negative- and positive-frequency parts of the spectral function:\be
A(\omega,k)=A_<(\omega,k)+A_>(\omega,k).
\ee
The Lehmann representation reads:\bea
A_<(\omega,k)&=&2\pi \sum_{n}|\langle \psi_n|c^{\phantom\dagger}_k|\psi_0\rangle|^2\delta(\omega+E_n-E_0),\\
A_>(\omega,k)&=&2\pi \sum_{n}|\langle \psi_n|c^{\dagger}_k|\psi_0\rangle|^2\delta(\omega-E_n+E_0),
\eea 
where $c_k$ annihilates a fermion with momentum  $k$ and $|\psi_n\rangle$ denotes an exact eigenstate of the Hamiltonian with energy $E_n$.
 Let us focus on the negative-frequency part  for $k<k_F$. For free fermions, we have   $A^{(0)}_<(\omega,k)=\delta(\omega
-\epsilon_0(k))$, where $\epsilon_0(k)<0$  is the energy  of the particle annihilated  below the Fermi surface.
When  weak interactions are turned on, the  renormalized
fermion dispersion  $\epsilon(k)$ becomes a threshold of
the support of $A_<(\omega,k)$, such that a    power-law singularity
develops for $\omega<\epsilon(k)$.   To describe the edge singularity for fixed  
$k<k_F$, we go back to the  mode expansion in Eq. (\ref{modeexpansion})
and generalize it to include three patches of momentum: 
\be 
c_j\to  \sqrt{a_0}[e^{ik_F x}r(x)+e^{-ik_F x}l(x)+e^{ikx}   \chi^\dagger(x)],\label{expandwithd}
\ee
Here the ``impurity field''  $\chi^\dagger(x)$  creates a hole in a state
with momentum close to $k$, within a subband of width $\Lambda\ll
k_F-k$. The low-energy Fermi fields $r(x)$ and $l(x)$ are also defined
with a cutoff of order $\Lambda$. In the case $q=k_F-k\ll k_F$, the
field $l(x)$ can be regarded as the projection of $L(x)$ into a
narrower subband, while $R(x)$ is split into two separate subbands
corresponding to the low-energy mode $r(x)$ and the ``high-energy''
mode $\chi(x)$. Restricting the energy window to the vicinity of the threshold with a single impurity, we now have to calculate the propagator of $\chi(x)$:\bea
G_{\text{ret},<}(\omega,k)&\approx& -i \int_0^{\infty} dt\, e^{i\omega t} \int_{-\infty}^{\infty} dx\nonumber\\
&&\times\langle \psi_0|\chi(0,0)\chi^\dagger(t,x)|\psi_0\rangle. \label{propchi}
\eea

At weak coupling,  i.e. as long as the four-fermion interaction
strength $V$ is small, we can  substitute  Eq. (\ref{expandwithd})
into Eq. (\ref{tVmodel}) and bosonize the low-energy fields to derive
an effective Hamiltonian for the single hole coupled to low-energy
collective modes. The result is the mobile impurity model
\cite{work5}
\bea 
\mc H_{\rm imp}&=&\int dx
\Big\{\frac{v}{16\pi}\Big[K(\partial_x\Theta)^2+\frac{ 1}K(\partial_x\Phi)^2
 \Big]\nonumber\\
 &&+\chi^\dagger(\varepsilon-iu\partial_x)\chi\nonumber\\ 
&&+\chi^\dagger \chi [f(q)\partial_x\varphi+\bar f(q)\partial_x\bar\varphi]\Big\},\label{impmodelspinless}
\eea
where $\varepsilon\equiv -\epsilon(k)>0$ is the energy of the ``deep hole'' excitation, $u=\frac{d\epsilon}{dk}$ is the velocity obtained by linearizing the dispersion around the centre of the impurity subband, and   $f(q)$ and $\bar f(q)$ are momentum-dependent impurity-boson couplings  of order $V$.  
The calculation of the Green's function in Eq. (\ref{Greensfunction})
is made possible by performing a unitary transformation that decouples
the impurity from the low-energy modes:\be
U=e^{-i\int_{-\infty}^{\infty}dx\,[\gamma\varphi(x)+\bar\gamma \bar \varphi(x)]\chi^\dagger(x) \chi(x)}.
\ee 
The  bosonic fields transform as\bea
\varphi^\circ(x)&=&U\varphi(x)U^\dagger=\varphi(x)-2\pi \gamma C(x),\nonumber\\
\varphi^\circ(x)&=&U\bar \varphi(x)U^\dagger=\varphi(x)+2\pi \bar\gamma C(x),
\eea
where \be
C(x)=\int_{-\infty}^{\infty} dy \,\text{sgn}(x-y)\chi^\dagger (y)\chi(y).
\ee
The transformed impurity field is\bea
d(x)&=&U \chi(x)U^\dagger\nonumber\\
&=&\chi(x) e^{i[\gamma\varphi(x)+\bar \gamma\bar\varphi(x)]}e^{-i\pi (\gamma^2-\bar \gamma^2)C(x)}.
\eea
Note that the impurity density is invariant under the unitary transformation, i.e. $\chi^\dagger(x)\chi(x)=d^\dagger(x)d(x)$.

We choose the  parameters $\gamma,\bar\gamma$ as the solution of\bea
\left(\begin{array}{c}f\\\bar f\end{array}\right)=\left(\begin{array}{cc}v_+-u&v_-\\ -v_-&-v_+-u\end{array}\right)\left(\begin{array}{c}\gamma\\ \bar\gamma\end{array}\right),
\eea
where
\be
v_\pm=\frac{v}2\left(K\pm \frac1K\right).
\ee
With this choice, the Hamiltonian becomes noninteracting:
\bea 
H_{\rm imp}&=&\int dx
\Big\{\frac{v}{16\pi}\Big[K(\partial_x\Theta^\circ)^2+\frac{ 1}K(\partial_x\Phi^\circ)^2
 \Big]\nonumber\\
 &&+d^\dagger(\varepsilon-iu\partial_x)d+\dots\Big\},\eea
 where $\dots$ stands for irrelevant operators which are neglected in the impurity model (since they only introduce subleading corrections to edge singularities). 
On the other hand, the expression in Eq. (\ref{propchi}) now becomes \bea
 G_{\text{ret,<}}(\omega,k)&\approx & -i\int_{0}^{+\infty}dt\, e^{i\omega t} \int_{-\infty}^\infty dx \langle d^{\phantom\dagger}(0,0)  d^{\dagger}(t,x)\rangle_0\nonumber\\
 &&\times\langle F(0,0)F^\dagger(t,x)\rangle_0,\label{shiftedG}
\eea
where $\langle\,\rangle_0$ denotes the expectation value in the noninteracting ground state $|\tilde\psi_0\rangle =U|\psi_0\rangle$ and $F(x)$ is the string operator \be
F(x)=e ^{i[\gamma \varphi^\circ(x)+\bar \gamma \bar\varphi^\circ(x)]}.\label{defstring}
\ee
The correlation function in Eq. (\ref{shiftedG}) can then be calculated by standard methods \cite{SIG3}. The important point is that the scaling dimension of the operator $ F(x)$ changes continuously as a function of $\gamma,\bar\gamma$. As a result, the effective impurity model predicts a power-law singularity in the spectral function, \be
A_<(\omega,k)\sim \theta(\epsilon(k)-\omega)|\epsilon(k)-\omega|^{-1+2(\gamma^2+\bar\gamma^2)}.\label{expspinless}\ee


Importantly, the  impurity mode $\chi(x)$  in   Eq. \eqref{impmodelspinless} carries  charge $q=1$  because it is defined from the original fermion $c_j$ at the noninteracting point. This is the particle that can be identified with an elementary excitation in the  Bethe ansatz solution for the integrable model. From the exact S-matrix it is known that interactions between these elementary excitations increases  as $V$ increases. Particularly at the SU(2) point, $V=2$, the elementary excitations are rather strongly interacting.  By contrast, the transformed impurity operator $d(x)=  \chi(x) F(x)$  carries a fractional charge that depends on the interaction strength, since the string  $F(x)$  in general   does not commute with the charge operator in Eq. (\ref{charge}). 

In the low-energy limit, an alternative approach to obtain the edge singularity in the spectral function was put forward  in Ref. \onlinecite{work8}.  In this approach one starts by introducing fermionic quasiparticles that are asymptotically free at low energies. In our notation, the idea is to define   chiral bosons $\phi,\bar\phi$ by\bea
\Phi(x)&=&\sqrt{K}(\phi+\bar\phi),\\
\Theta(x)&=&\frac1{\sqrt{K}}(\phi-\bar\phi).
\eea
In terms of these, the Luttinger model (\ref{HLL}) reads\be
\mc H_{\text{LL}}(x)=\frac{v}{8\pi}[(\partial_x\phi)^2+(\partial_x\bar\phi)^2].\label{decoupledphis}
\ee
The quasiparticles $\tilde R(x)$ and $\tilde L(x)$ are defined by\bea
\tilde R(x)&=&\frac{\eta}{\sqrt{2\pi}}e^{-\frac{i}{\sqrt2}\phi(x)},\label{quasiR}\\
\tilde L(x)&=&\frac{\bar\eta}{\sqrt{2\pi}}e^{\frac{i}{\sqrt2}\bar \phi(x)}.\label{quasiL}
\eea
The commutator  with the charge operator yields\be
[ q,\tilde R^\dagger(x) ]=\sqrt{K}\tilde R^\dagger(x),\quad  [q, \tilde L^\dagger(x) ]=\sqrt{K}\tilde L^\dagger(x).
\ee
Thus, the quasiparticles carry   charge $\sqrt{K}$. On the other hand, this refermionization procedure removes the marginal interaction between the quasiparticles since the chiral modes are decoupled in Eq. (\ref{decoupledphis}) \cite{work8,rozhkov}. The leading interaction is  then represented by the irrelevant operator $g_3$ in Eq. (\ref{irretas}), which can be neglected as a first approximation in the low-energy limit. 
 The relation between the original right-moving  fermion and the new quasiparticle is  \be
\tilde R(x)=  R(x) F_0(x),\ee
where $F_0(x)$ is the limit $q=k_F-k\to 0$ of the string operator    in Eq. (\ref{defstring}).  At this point a noninteracting  impurity mode $\tilde\chi(x)$ can be introduced by projecting  the free field $\tilde R(x)$ into low-energy and high-energy subbands. This leads to a universal result for the  exponent in the vicinity of threshold, $|\omega-\epsilon(k)|\ll k^2/\tilde m$,   which corresponds to Eq. (\ref{expspinless}) with parameters $\gamma=\frac1{\sqrt2}\left(1-\frac{1}{2\sqrt{K}}-\frac{\sqrt{K}}{2}\right)$ and $\bar \gamma=\frac1{\sqrt2}\left(\frac{1}{2\sqrt{K}}-\frac{\sqrt{K}}{2}\right)$. This result   differs from the prediction of the linear theory\cite{work8}.

In summary, there are   two possible   paths towards calculating edge
exponents in the nLL theory for spinless fermions: (i)  starting with
free fermions, one  defines low-energy and impurity subbands, and then
turns on interactions between the elementary excitations in the
impurity model; after that, the interaction with the impurity  is
removed by a unitary transformation, which introduces the string
operator in the correlation function; or (ii) starting from the
Luttinger model for interacting fermions, one refermionizes into
weakly interacting quasiparticles, which differ from the original
fermions by a string operator,  and then projects the quasiparticles
into low-energy and impurity subbands. The projection onto the
impurity model  is  well controlled in both paths  because the   model
of interacting spinless fermions  is smoothly connected with the free
model; i.e., the parameters $\gamma,\bar\gamma$, which quantify   the
scattering between high-energy and low-energy particles, vanish
continuously as $V\to0$. However, it is important to emphasize the
difference between the original fermions, which carry unit charge of
the U(1) symmetry, and the quasiparticles with fractional
charge. While the latter are always weakly interacting in the
low-energy limit, the fermions that carry the correct quantum numbers
become strongly interacting even at low energies as $V$ increases.

Once the low-energy, weak-coupling regime is well understood, the
impurity model of nLL  theory can be extended phenomenologically to
high energies, strong interactions and thresholds with more than one
impurity, as has indeed been done successfully for spinless fermions
\cite{SIG3}.

\section{Spinful Fermions}
\begin{figure}[ht]
\begin{center}
\epsfxsize=0.45\textwidth
\epsfbox{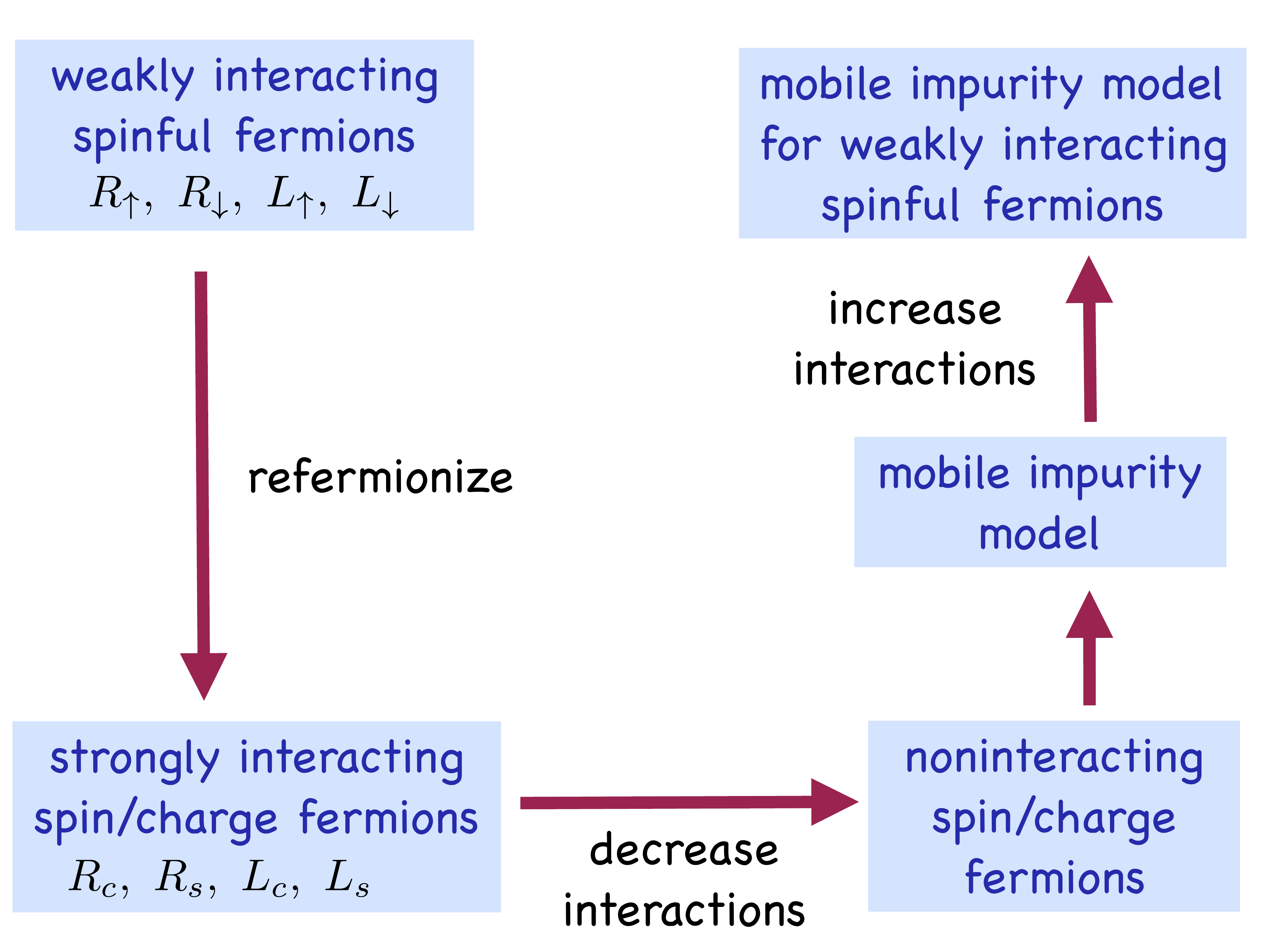}
\end{center}
\caption{Strategy map for calculating edge exponents in nonlinear
Luttinger liquid theory for spinful fermions in the low-energy
limit. See the main text for details.
} 
\label{fig:strong}
\end{figure}

As we have seen above, the spinless case is most easily understood by
considering the vicinity of noninteracting fermions. The situation is
very different in the spinful case. In order to understand this point
in some detail, let us consider the particular example of the 1D
Hubbard model 
\be
H_{\textrm{Hub}}=-t\sum_{j=1}^L\sum_{\sigma}(c_{j,\sigma}^\dagger 
c_{j+1,\sigma}^{\phantom\dagger}+{\rm h.c.})+U \sum_{j}n_{j,\uparrow}n_{j,\downarrow},
\label{HHubb}
\ee
where $c_{j,\sigma}$ annihilates a fermion with spin
$\sigma=\uparrow,\downarrow$ at site $j$,
$n_{j,\sigma}=c_{j,\sigma}^\dagger c_{j,\sigma}^{\phantom\dagger}$ is
the  number operator, and $U\geq 0$ is the strength of the  on-site
repulsion. We work at fixed density below half-filling with zero
magnetization, $N_{\uparrow}=N_{\downarrow}<L/2$. In this case the
model has a global U(1)$\otimes$SU(2) symmetry. Let us focus first on
the limit of weak interactions $U/t\ll 1$ and low energies. It is well
known \cite{hubbardbook} that the low-energy degrees of freedom are
collective spin and charge modes respectively, i.e.
\be
H_{\rm Hub}\longrightarrow {\cal H}_{\rm charge}+{\cal H}_{\rm spin}+\ldots
\ee
where the dots denote additional terms that are irrelevant in the
renormalization group sense. Crucially, as $H_{\rm Hub}$ is spin
rotationally symmetric, ${\cal H}_{\rm spin}$ must exhibit a spin
SU(2) symmetry. In order to parallel our analysis in the spinless case
we wish to express ${\cal H}_{\rm spin}$ in terms of fermionic fields
carrying spin quantum numbers $\pm 1/2$. This is possible
in an SU(2)-symmetric way only if the fermions are \emph{strongly 
interacting}, i.e., the situation is similar to the $V=2$ case for spinless
fermions. In the charge sector the situation is analogous unless we
work at very low electron densities. In order to generalize the mobile
impurity model construction reviewed in section \ref{sec:spinless} to
the spinful case, we therefore cannot work with weakly interacting 
spinful fermions, but require a model that gives rise to
noninteracting fermions describing the collective spin and charge
degrees of freedom. As such a model is not known, we proceed along the
lines sketched in Fig.~\ref{fig:strong}.
\begin{enumerate}
\item{} Starting with weakly interacting spinful fermions at low
energies, we derive the corresponding model of strongly interacting
spin and charge fermions. 
\item{} We then decrease the interactions in the spin and charge fermion
model, and derive a low-energy effective Hamiltonian in the vicinity
of the ``Luther-Emery point'' \cite{luther}  where the spin/charge fermions
become noninteracting.
\item{} Having completed this construction, we are in a position to
construct mobile impurity models by following the logic employed in
the spinless case.
\item{} Having constructed a suitable mobile impurity model, we may
calculate threshold exponents by standard methods.
\item{} Through an appropriate tuning of the parameters defining our
mobile impurity model, we may analyze the case of weakly interacting
SU(2)-invariant spinful fermions. This is analogous to the analysis of
strongly interacting spinless fermions with SU(2) symmetry based on a
mobile impurity model formulated at weak coupling.
\end{enumerate}
A key ingredient in our approach is our ``Luther-Emery model'' of
noninteracting holons and spinons. An obvious question is what such a
theory might look like in terms of interacting spinful fermions. We
address this issue in section \ref{sec:LEP}.

\subsection{Bosonization at weak coupling}
As our point of departure we choose a general extended Hubbard model
below half-filling, where we allow fairly general electron-electron
interactions in addition to \fr{HHubb}, provided that they are
invariant under the following symmetries:
\begin{itemize}
\item{} U(1)$\otimes$ U(1) transformations in the charge and spin
  sectors
$c_{j,\sigma}\rightarrow e^{i\alpha_\sigma}c_{j,\sigma}$, $\alpha_\sigma\in\mathbb{R}$;
\item{} spin reflection $c_{j,\up}\leftrightarrow c_{j,\down}$;
\item{} site parity $c_{j,\sigma}\rightarrow c_{-j,\sigma}$;
\item{} translations $c_{j,\sigma}\rightarrow c_{j+1,\sigma}$.
\end{itemize}
The kind of lattice model we have in mind is of the form
\bea
H&=&H_{\textrm{Hub}}+\sum_{r\geq 1}\sum_{j}V_rn_{j}n_{j+r}\nn
&&+\sum_{r\geq 1}\sum_{j}
\left(J_r{\bf S}_{j}\cdot{\bf S}_{j+r}+J^z_rS^z_{j}S^z_{j+r}\right),
\label{HEHubb}
\eea
where the coupling constants $\{V_r,J_r,J_r^z\}$ must be such that
the model remains in a spin-charge-separated quantum critical phase.
Lattice models of this type can be bosonized by standard methods
\cite{gogolin,giamarchi}. The generalization of Haldane's bosonization
formulas \cite{haldane} to the spinful case is
\bea
c_{j,\sigma}&\sim&\sqrt{a_0}
\sum_{n,m\in\mathbb{Z}}\Gamma^{(\sigma)}_{n,m}
A_{n,m}e^{ik_Fx(1-4n-2m)}\nonumber\\
&\times& e^{-\frac i2\varphi_c(x)-\frac i2 s_\sigma
  \varphi_s(x)}e^{i\frac{2n+m}{2}\Phi_c(x)}e^{-is_\sigma\frac{m}2\Phi_s(x)}.
\label{opexpansion}
\eea
Here $s_\up=1=-s_\down$, $a_0$ is a short-distance cutoff, $k_F=\pi
N_\uparrow/(La_0)$ is the Fermi momentum, $A_{n,m}$ are non-universal
amplitudes and $\Gamma_{n,m}^{(\sigma)}\equiv \eta_\sigma
(\eta_\sigma\bar\eta_\sigma\eta_{\bar\sigma}\bar\eta_{\bar\sigma})^{n}
(\eta_{\bar\sigma}\bar\eta_{\bar\sigma})^{m}$
are Klein factors (and we use notations where e.g. $\bar\up=\down$)
that ensure the correct anti-commutation relations. The bosonic fields
\bea
\Phi_{\alpha}(x)&=&\varphi_\alpha(x)+\bar\varphi_\alpha(x),\\
\Theta_\alpha(x)&=&\varphi_\alpha(x)-\bar\varphi_\alpha(x),
\eea
with $\alpha=c,s$, obey commutation relations  
\be
[\Phi_\alpha(x),\Theta_{\alpha^\prime}(x^\prime)]=4\pi
i\delta_{\alpha,\alpha^\prime}
\textrm{sgn}(x-x^\prime).
\ee
We note that in the CFT normalizations \fr{normalization} the
amplitudes $A_{n,m}$ are dimensionful, i.e. are proportional to
appropriate powers of the lattice spacing $a_0$.

In the spin-charge separated Luttinger liquid phase the low-energy
effective Hamiltonian for extended Hubbard models of the type \fr{HEHubb}
is
\bea
{\cal H}&=&\int dx\left[{\cal H}_{\rm LL}(x)+{\cal H}_{\rm irr}(x)\right] ,\\
{\cal H}_{\rm LL}(x)&=&
\sum_{\alpha=c,s}\frac{v_\alpha}{16\pi}\Big[K_\alpha(\partial_x\Theta_\alpha)^2
+\frac{1}{K_\alpha}(\partial_x\Phi_\alpha)^2\Big],\label{LLwithchargeandspin}\\
{\cal H}_{\rm
  irr}(x)&=&\lambda_1\cos\Phi_s+\lambda_2\partial_x\Phi_c
\cos\Phi_s\nn
&+&\sum_{\alpha=c,s}\lambda_{3,\alpha}^\pm\partial_x\Phi_c
\left[\big(\partial_x\Phi_\alpha\big)^2\pm
\big(\partial_x\Theta_\alpha\big)^2\right]\nn
&+&\lambda_4\partial_x\Phi_s\partial_x\Theta_s\partial_x\Theta_c
+\ldots
\label{luttinger} 
\eea
Here $v_\alpha$ are the velocities of the collective charge and
spin modes and $K_\alpha$ the corresponding Luttinger parameters.
The contributions ${\cal H}_{\rm irr}$ are irrelevant in the
renormalization group sense. A complete list of irrelevant operators
with scaling dimensions of at most four (for $K_s= 1$) is given in
Appendix~\ref{app:irrops}. The velocities $v_\alpha$ and Luttinger 
parameters $K_\alpha$ can be calculated exactly for the Hubbard
model, but Eq. (\ref{luttinger}) is generic for spinful Luttinger
liquids if we regard $v_\alpha$ and $K_\alpha$ as phenomenological
parameters. We note that, as a consequence of spin reflection symmetry,
marginal interactions coupling spin and charge such as 
\be
\partial_x\Phi_s(x)\partial_x\Phi_c(x)
\ee
are not allowed. Hence the collective degrees of freedom  at low energies are
described in terms of pure spin and pure charge modes, rather than
linear combinations thereof (which would be the case in presence of a
magnetic field, see e.g. Ref. \onlinecite{frahmkorepin2}).

\subsection{Refermionizing in terms of spin and charge fields}
The next step is to refermionize (\ref{luttinger}) in terms of spin
and charge fermion fields. In order to see how this should be done, we
consider the limit of vanishing interactions. Here the bosonization
formulas simplify to ($\sigma=\up,\down$)
\bea
c_\sigma&\rightarrow&\sqrt{a_0}\left[e^{ik_Fx}R_\sigma(x)+e^{-ik_Fx}L_\sigma(x)\right],\nn
R_\sigma(x)&\sim&\frac{\eta_\sigma}{\sqrt{2\pi}}\ e^{-\frac i2\varphi_c(x)-\frac
  i2s_\sigma\varphi_s(x)}\ ,\nn
L_\sigma(x)&\sim&\frac{\bar\eta_\sigma}{\sqrt{2\pi}}\ e^{\frac i2\bar\varphi_c(x)+\frac
  i2s_\sigma\bar\varphi_s(x)}\ ,
\eea
where $s_\up=-s_\down=1$. The idea is to decompose the right-moving
spin-up electron into a right-moving holon field $R_c$ and a
right-moving spinon field $R_s$ in the form
\be 
R_{\uparrow}(x)\sim R_c(x)e^{i\textrm{ (charge string)}}R_s(x)e^{i\textrm{ (spin string)}}.\label{defineholonspinon}
\ee
In a pure Luttinger liquid there are infinitely many acceptable choices
for $R_{\alpha}$ and string operators in Eq. \eqref{defineholonspinon}. 
In Refs. \onlinecite{SIG1,SIG2}, the fermions  $R_\alpha$ are chosen so as to
have scaling dimension $1/2$, in analogy with the procedure in the
spinless case, see Eqs. (\ref{quasiR}) and (\ref{quasiL}). This choice is
such that the particles 
are asymptotically free at low energies. However, they then carry
fractional spin and charge quantum numbers. Such a choice is not the
most natural one for our purposes: it is known that the scaling limit
of the Hubbard model is given by the U(1) Thirring model (SU(2) at
half-filling), see e.g. Ref.~\onlinecite{gogolin}. The U(1) Thirring model is
integrable, and the elementary excitations are known to be strongly
interacting fermionic spinless holons and neutral spinons (with a
known S-matrix) carrying charge $\mp e$ and spin $\pm 1/2$
respectively. The principle guiding our construction is that charge
and spin fermions created by $R_c^\dagger $ and $R_s^\dagger$ should
carry the same quantum numbers as the elementary holon and spin
excitations. The U(1) charges corresponding to these quantum numbers are
\be
q_{\alpha}=\int_{-\infty}^{+\infty} dx\ Q_\alpha(x),
\ee
where 
\bea
Q_c(x)&=&\sum_{\sigma=\up,\down}\left[R^\dagger_\sigma(x)R_\sigma(x)+L^\dagger_\sigma(x)L_\sigma(x)\right],\nn
Q_s(x)&=&\sum_{\sigma=\up,\down}{s_\sigma}
\left[R^\dagger_\sigma(x)R_\sigma(x)+L^\dagger_\sigma(x)L_\sigma(x)\right].
\eea
As usual these expressions are to be understood in terms of a standard
point splitting and normal ordering prescription. We now require
\bea
Q_c(x)=R^\dagger_c(x)R_c(x)+L^\dagger_c(x)L_c(x)\ ,\nn
Q_s(x)=R^\dagger_s(x)R_s(x)+L^\dagger_s(x)L_s(x),
\eea
which ensure that the spin and charge fermions have the desired U(1)
charges 
\be
[q_\alpha,R^\dagger_\alpha(x)]=R^\dagger_\alpha(x),\
[q_\alpha,L^\dagger_\alpha(x)]=L^\dagger_\alpha(x).
\ee
Our refermionization prescription then reads
\bea
R_\up(x)&\sim& \eta_\up\ {\cal O}_c(x)\ {\cal O}_s(x)\ ,\nn
R_\down(x)&\sim& \eta_\down\ {\cal O}_c(x)\ {\cal O}^\dagger_s(x)\ ,\nn
{\cal O}_\alpha(x)&\sim& R_\alpha(x)\ e^{-\frac{i\pi}{2}\int_{-\infty}^x
  dx'\ Q_\alpha(x')}\ .
\label{ourfermions}
\eea
Analogous relations hold for left-moving fermions.
One issue that arises here is that $n_s(x)=\int_{-\infty}^xdx' Q_s(x')$ is
the number of spinons on the interval $[-\infty,x]$, and therefore
string operators of the form $\exp\big[i\alpha n_s(x)\big]$ should be
$2\pi$-periodic functions of $\alpha$. When bosonizing string
operators naively this periodicity is lost. A simple way of dealing
with this issue is via the replacement \cite{AEM}
\be
\exp\big[i\alpha n_s(x)\big]\longrightarrow
\sum_m\exp\big[i(2\pi m+\alpha) n_s(x)\big].
\label{periodicstring}
\ee
The operators ${\cal O}_\alpha(x)$ fulfill braiding relations for
$x\neq y$:
\be
{\cal O}_\alpha(x){\cal O}_\alpha(y)=
e^{-\frac{i\pi}{2}{\rm sgn}(x-y)}{\cal O}_\alpha(y){\cal O}_\alpha(x).
\ee
The low-energy effective Hamiltonian \fr{luttinger} is expressed in
terms of our fermionic charge and spin fields as 
\bea 
{\cal H}&=&  \int dx \left[{\cal H}_c(x)+{\cal H}_s(x)+
{\cal H}_{cs}(x)\right],\nn
{\cal H}_c&=&R^\dagger_c (-iv_c^\prime \partial_x-\eta\partial_x^2)
R^{\phantom\dagger}_c+L^\dagger_c (iv_c^\prime
\partial_x-\eta\partial_x^2) L^{\phantom\dagger}_c\nn
&&+g_{c,0} R^\dagger_cR^{\phantom\dagger}_cL^\dagger_cL_c+\ldots,\nn
{\cal H}_s&=&R^\dagger_s(-iv_s^\prime \partial_x+i\zeta
\partial_x^3)R^{\phantom\dagger}_s+L^\dagger_s(iv_s^\prime
\partial_x-i\zeta \partial_x^3)L^{\phantom\dagger}_s\nn
&&+g_{s,0} R^\dagger_sR^{\phantom\dagger}_sL^\dagger_sL^{\phantom\dagger}_s
+ g_{s,1} (R^\dagger_s
L^{\phantom\dagger}_s+L^\dagger_sR^{\phantom\dagger}_s)+\dots,\nn
{\cal H}_{cs}&=&
g_1 (R^\dagger_c
R^{\phantom\dagger}_c+L^\dagger_cL^{\phantom\dagger}_c)(R^\dagger_s
L^{\phantom\dagger}_s+{\rm h.c.})+\dots .\label{Hcs}
\eea
A crucial feature of this expression is that the coupling constants of
the marginal interactions,
\be
g_{\alpha,0}\sim 2\pi v_\alpha\left(\frac{1}{4K_\alpha}-K_\alpha\right),
\label{galpha}
\ee
  are ${\cal O}(1)$ at weak coupling $K_\alpha\to 1$. Moreover
$g_s$ is always large as long as the spin SU(2) symmetry is
unbroken, as in this case the Luttinger parameter is fixed at
$K_s=1$. This implies that the spin and charge fermions are
\emph{strongly interacting}. This is consistent 
with known results for the exact S-matrix of the Hubbard model
\cite{EK94a,EK94b,Andrei}. Moreover, the spin sector of \fr{Hcs} 
describes a massive Thirring model perturbed by irrelevant operators,
which is precisely what one would expect on the basis of the known
S-matrices for the Hubbard model \cite{coleman}.

\subsection{Bosonic representation of charge and spin fermions}
\label{app:boso}
Our spin and charge fermion fields can be bosonized by standard
methods. Introducing chiral charge and spin ($\alpha=c,s$) Bose fields
$\varphi_\alpha$, $\bar\varphi_\alpha$, and ignoring higher harmonics,
we have
\be
R_\alpha(x)\sim \frac{\eta_\alpha}{\sqrt{2\pi}}
e^{-\frac{i}{\sqrt2}\varphi_\alpha^*(x)} ,\
L_\alpha(x)\sim \frac{\bar\eta_\alpha}{\sqrt{2\pi}}
e^{\frac{i}{\sqrt2}\bar\varphi_\alpha^*(x)}\ ,\label{spinlessRalpha}
\ee
where $\eta_\alpha$, $\bar\eta_\alpha$ are Klein factors fulfilling
anticommutation relations
$\{\eta_\alpha,\eta_\beta\}=2\delta_{\alpha,\beta}
=\{\bar\eta_\alpha,\bar\eta_\beta\}$, $\{\bar\eta_\alpha,\eta_\beta\}=0$.
Bosonizing the spin and charge fermions leads to the following
expressions for  the original right- and left-moving spinful fermions
\bea
R_\uparrow(x)&\propto& \prod_{\alpha=c,s}e^{-\frac{i}{\sqrt2}
  \varphi_\alpha^*(x)+\frac{i}{4\sqrt{2}}
\Phi^*_\alpha(x)}\ ,\nn
L_\uparrow(x)&\propto& \prod_{\alpha=c,s}e^{\frac{i}{\sqrt2}
  \bar\varphi_\alpha^*(x)-\frac{i}{4\sqrt{2}}
\Phi^*_\alpha(x)}\ .
\label{boso_sc}
\eea
The new Bose fields $\varphi^*_{\alpha}$, $\bar\varphi^*_{\alpha}$ are
related to the usual spin and charge bosons \fr{luttinger} by a
canonical transformation  
\be
\Phi_\alpha=\frac{\Phi_\alpha^*}{\sqrt{2}}\ ,\quad
\Theta_\alpha=\sqrt{2}\Theta_\alpha^*\ .
\label{old_new}
\ee
Given \fr{old_new}, it is straightforward to rewrite \fr{luttinger} in
terms of the new Bose fields
\bea
H_{b}&=&\int dx\,\left\{\sum_{\alpha}\frac{v_\alpha^\prime}{16\pi}[(\partial_x\Theta_\alpha^*)^2+(\partial_x\Phi_\alpha^*)^2]\right.\nonumber\\
&&+\sum_{\alpha} \lambda_\alpha[(\partial_x\Theta_\alpha^*)^2-(\partial_x\Phi_\alpha^*)^2]+\lambda_1\cos(\Phi^*_s/\sqrt2)\nonumber\\
&&\left.+\frac{\lambda_2}{\sqrt2}
\partial_x\Phi_c^*\cos(\Phi^*_s/\sqrt2)+\dots\right\},\label{interactingbosons}
\eea
where
$v_{\alpha}^\prime=v_\alpha\left(K_\alpha+\frac{1}{4K_\alpha}\right)$
and
$\lambda_\alpha=\frac{v_\alpha}{16\pi}\left(K_\alpha-\frac{1}{4K_\alpha}\right)$.

\section{Luther-Emery (LE) point for spin and charge}
\label{sec:LEP_FT}
A particular case of the family of Hamiltonians \fr{Hcs} describes
a free theory of \emph{non-interacting} gapless fermionic spinons and
holons. This LE point for both spin and charge corresponds to
\bea 
{H}_{\rm LE}&=& \int dx\Big\lbrack
R^\dagger_c (-iv_c^\prime \partial_x-\eta\partial_x^2+\dots)
R^{\phantom\dagger}_c\nn
&&+L^\dagger_c (iv_c^\prime \partial_x-\eta\partial_x^2+\dots)
L^{\phantom\dagger}_c\nn
&&+R^\dagger_s(-iv_s^\prime \partial_x+i\zeta
\partial_x^3+\dots)R^{\phantom\dagger}_s\nn
&&+L^\dagger_s(iv_s^\prime \partial_x-i\zeta
\partial_x^3+\dots)L^{\phantom\dagger}_s\Big\rbrack.
\label{HLE}
\eea
Here we have included the quadratic (cubic) term in the holon
(spinon) dispersion to emphasize the nonlinearity.
In order to realize a Hamiltonian of this form fine-tuning a number of
couplings is required, as can be seen by analyzing the stability of
\fr{HLE} to perturbations.
\subsection{Stability of the LE point}
\label{ssec:stability}
An obvious question is to what extent the LE point is stable. The most
important perturbations to \fr{HLE} are
\bea 
H_{\rm pert}&=& \int dx \Big\lbrack 
g_{c,0} R^\dagger_cR_c L^{\phantom\dagger}_cL_c\nn
&&+g_{s,1} (R^\dagger_sL^{\phantom\dagger}_s+L^\dagger_sR^{\phantom\dagger}_s)
+ g_{s,0} R^\dagger_sR_s L^{\phantom\dagger}_sL_s\nn
&&+g_1  (R^\dagger_c
R^{\phantom\dagger}_c+L^\dagger_cL^{\phantom\dagger}_c)(R^\dagger_s
L^{\phantom\dagger}_s+{\rm h.c.})\Big\rbrack.
\label{perturbations}
\eea
In addition to \fr{perturbations} there are other, less relevant
perturbations. A list of the ones with scaling dimension below four is
given in Appendix~\ref{app:irrLE}. 
The $g_{s,1}$ term in (\ref{perturbations}) is
recognized as a mass term for spinons, and is the only strongly
relevant perturbation. This implies that spinons are generically
gapped, and in order to reach a LE point with gapless spinons
\emph{fine tuning} $g_{s,1}=0$ is necessary. Assuming that this is possible,
we are left with three perturbing operators of scaling dimension $2$.
While the $g_{s,0}$ and $g_{c,0}$ terms are scalar, the $g_1$ term
carries non-zero Lorentz spin. In order to assess the stability of the
LE point to these perturbations, we have carried out a renormalization
group analysis. In principle we need to work with different cut-offs
for the charge and spin degrees of freedom. However, at one-loop
logarithmic divergences are encountered only in the spin sector. We
obtain RG equations of the form 
\begin{eqnarray}
\frac{dg_1}{d\ell}&=&\frac{1}{4\pi\upsilon_s}g_1 g_{s,0},\quad 
\frac{d\upsilon_c}{d\ell}=-\frac{1}{2\pi^2\upsilon_s}g_1^2, \label{RGeq1} \\
\frac{d g_{c,0}}{d\ell}&=&-\frac{1}{\pi\upsilon_s}g_1^2, \quad \frac{d
  g_{s,0}}{d\ell}=\frac{d\upsilon_s}{d\ell}=0,
\label{RGeq2}
\end{eqnarray}
where $\ell=\ln(L/a_0)$ and $a_0$ and $L$ are short and long distance
cutoffs respectively. The RG equations are easily integrated
\bea
g_{s,0}(\ell)&=&g_{s,0}(\ell_0)\ ,\quad
g_{1}(\ell)=g_1(\ell_0)e^{\frac{g_{s,0}(\ell_0)}{4\pi
    v_s(\ell_0)}(\ell-\ell_0)}\ ,\nn
g_{c,0}(\ell)&=&g_{c,0}(\ell_0)-\frac{2g_1^2(\ell_0)}{g_{s,0}(\ell_0)}
\left[e^{\frac{g_{s,0}(\ell_0)}{2\pi
    v_s(\ell_0)}(\ell-\ell_0)}-1\right],
\eea
and imply the following:
\begin{enumerate}
\item{} The spinon mass term is not produced under the RG flow
if the bare coupling is initially set to zero. We have checked that
this remains true at two loops. However, we cannot rule out that
$g_{s,1}$ may be generated at higher orders and it is possible that
setting it to zero requires fine tuning an infinite number of
parameters in a lattice model.
\item{} The coupling $g_{s,0}$ does not flow under the RG. This
remains true at two-loop order. Hence, to this order, $g_{s,0}$ needs
to be fine-tuned to zero in order to reach the LE point.
\item{} If the initial value $g_{s,0}(\ell_0)<0$, the coupling
  $g_1(\ell)$ flows to zero under the RG, while $g_{c,0}(\ell)$ flows to a
  constant value.
\end{enumerate}

\subsection{Threshold singularities in the single electron spectral
  function} 
Given the low-energy Hamiltonian at the LE point \fr{HLE}, we are now
in a position to derive a mobile impurity model, valid a priori at low
energies. The usual continuity arguments suggest that the restriction
to low energies can be relaxed and the model applied to
energies of the order of the lattice scale $t$. Let us focus on the
mobile impurity model relevant for analyzing the threshold behaviour
in the single-electron spectral function 
\bea
A(\omega,k)&=&-\frac{1}{\pi}\ {\rm Im}\ G_{\rm ret}(\omega,k),\nn
G_{\rm ret}(\omega,k)&=&-i\int_0^\infty dt\ e^{i\omega t}\sum_l e^{-ikla_0}\nn
&&\quad\times\ \langle \psi_0|\{c_{j+l,\sigma}(t),\ c^\dagger_{j,\sigma}\}|\psi_0\rangle,
\label{specfun}
\eea
where $|\psi_0\rangle$ is the ground state. 

\begin{figure}[ht]
\begin{center}
\epsfxsize=0.42\textwidth
\epsfbox{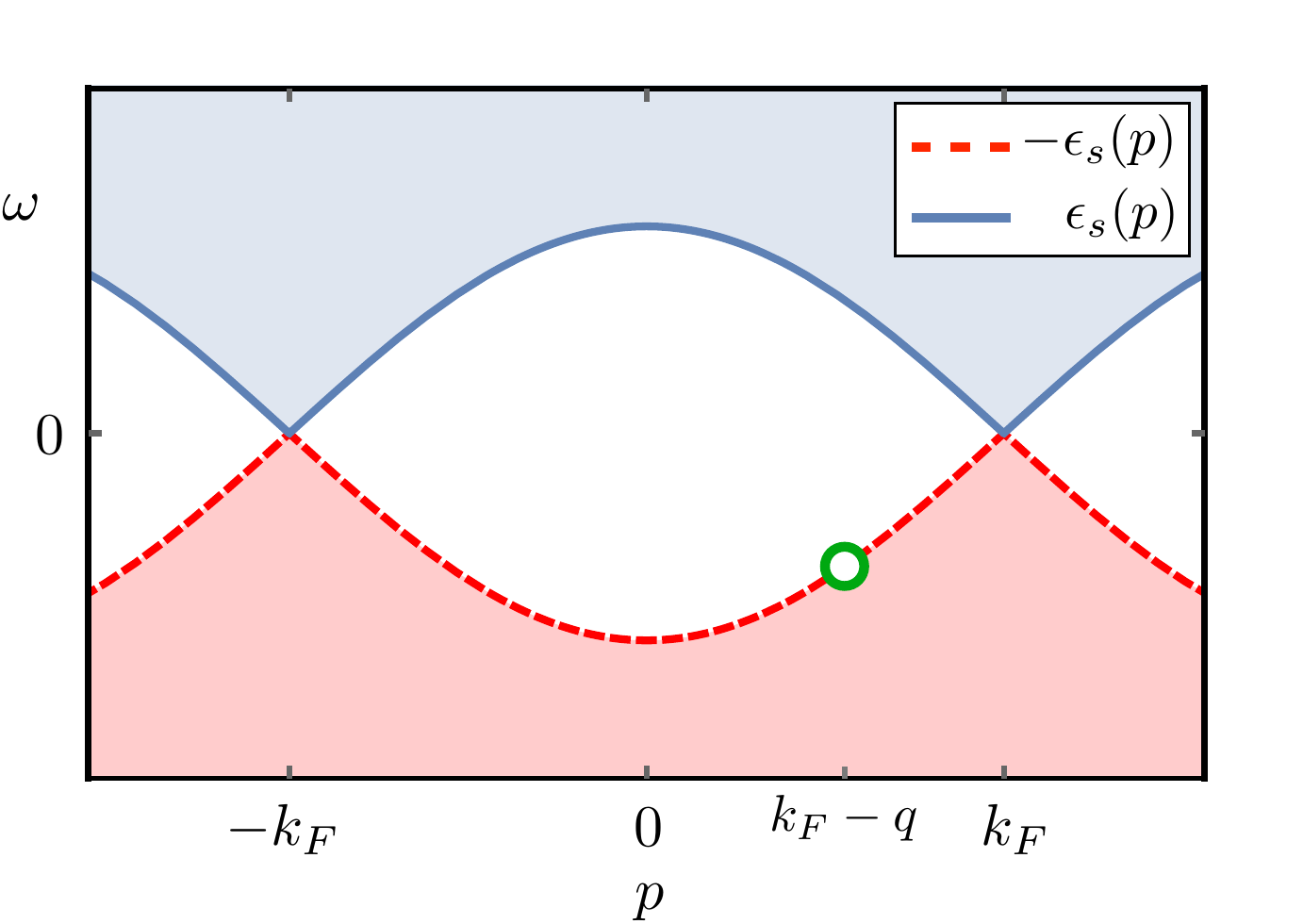}
\end{center}
\caption{Support in the energy/momentum plane of excitations with the
quantum numbers of an electron or hole. For commensurate band fillings
there is an absolute threshold that we take to follow the
spinon/anti-spinon dispersions. Above the threshold the
single-particle spectral function is singular, and we aim
to determine the threshold exponent of the negative-frequency part at a
momentum $k_F-q$ (green circle).}  
\label{fig:spinonthres}
\end{figure}
For commensurate band fillings the spectral function has a threshold at
low energies. To be specific, we will consider the case $v'_s<v'_c$, in
which case the threshold corresponds to exciting a single high-energy
spinon, while (anti)holon excitations have vanishing energy. The
corresponding kinematics is sketched in Fig.~\ref{fig:spinonthres}. The
negative frequency part of the spectral function at fixed momentum
transfer $k$ exhibits a threshold singularity
\be
 A(\omega,k)=
\begin{cases}
0 &{\rm if\ } 0>\omega>-\eps_s(k)\\
A_0|\omega+\eps_s(k)|^{\mu} & {\rm
  if\ }\omega\rightarrow-\eps_s(k)\ .
\end{cases}
\ee
Here $\epsilon_s(k)$ denotes the spinon dispersion.
\subsection{Threshold exponent at the LE point
\label{sec:LEexponent}}
Let us focus on momentum transfers $k_F-q$, where we take
$0<q\ll k_F$. Using the decomposition
\be
c_\sigma\sim\sqrt{a_0}\left[e^{ik_Fx}R_\sigma(x)+e^{-ik_Fx}L_\sigma(x)\right],
\ee
we see that the relevant field theory correlator is
\bea
A(\omega,k_F-q)&\sim&-i\int_0^\infty dt\int_{-\infty}^\infty dx\ e^{i\omega t+iqx}\nn
&&\times
\langle\psi_0|\lbrace
R_\uparrow(t,x),R_\up^\dagger(0,0)\rbrace|\psi_0\rangle\nn
&=&A_<(\omega,k_F-q)+A_>(\omega,k_F-q).
\eea
Here $A_>$ and $A_<$ are respectively the positive and negative
frequency parts of the spectral function. Using \fr{ourfermions} we
arrive at the following expression for the latter
\bea
A_<(\omega,k_F-q)&\sim&-i\int_0^\infty dt\int_{-\infty}^\infty dx\
e^{i\omega t+iqx}\nn
&\times&
\prod_{\alpha=c,s}\langle\psi_0|{\cal O}_\alpha^\dagger(0,0){\cal O}_\alpha(t,x)
|\psi_0\rangle\ .
\label{holepart}
\eea 
At the LE point we are dealing with a free fermion theory. Hence
correlation functions of the kind required in \fr{holepart} can be
expressed as Fredholm determinants \cite{ZCG}, but we do not follow
this route here. Instead, we construct a mobile impurity model and use
it to extract the threshold exponent.

At the LE point spin and charge degrees are perfectly separated.
As a consequence it is possible to construct a basis of energy
eigenstates in the form
\be
|n_c\rangle\otimes|n_s\rangle,
\ee
where $n_{c,s}$ are appropriate charge and spin quantum numbers. The
correlators required in \fr{holepart} then have Lehmann
representations of the form
\bea
\sum_{n_\alpha}e^{iE_{n_\alpha}t-iP_{n_\alpha}x}
|\langle n_\alpha|{\cal O}_\alpha(0,0)|\psi_0\rangle|^2\ ,\
\alpha=c,s.
\label{Lehmannrep}
\eea
The threshold singularity arises from excitations involving a single
high-energy spinon with momentum $q$ plus low-energy excitations in
the charge and spin sectors sector. This means that the charge part of
\fr{holepart} can be calculated using bosonization. The bosonization
identities \fr{boso_sc} imply that
\be
{\cal O}_c(x)\propto e^{-\frac{i}{\sqrt{2}}\varphi_c^*(x)}
e^{\frac{i}{4\sqrt{2}}\Phi_c^*(x)}.
\label{Oc}
\ee
At the LE point the coupling constants $\lambda_\alpha$, $\lambda_1$, $\lambda_2$ in
\fr{interactingbosons} vanish, and a simple calculation gives
\be
\langle {\cal O}_c^\dagger(0,0){\cal O}_c(t,x)
\rangle\propto
(v^\prime_ct-x)^{-\frac{1}{2}}\ (x^2-{v_c'}^2t^2)^{-\frac{1}{16}}.
\label{OcOc}
\ee
In order to work out the contribution from the spin part we follow
Ref.~\onlinecite{karimi}. We decompose the spin fermions into
low-energy and mobile impurity parts
\bea
R_s(x)&\sim& r_s(x)+e^{-iqx}\chi^\dagger_s(x)\ ,\nn
L_s(x)&\sim&  l_s(x),
\label{decomp}
\eea
where $\chi_s^\dagger$ creates a hole in the spinon band.  In terms of
momentum modes
\be
R_{s}(x)\sim \int \frac{dp}{2\pi}\ e^{ipx}R_s(p),
\ee
this projection corresponds to
\bea
r_s(x)&\sim& \int_{-\Lambda'}^{\Lambda'}
\frac{dp}{2\pi}\ e^{ipx}R_s(p),\nn
\chi_s^\dagger(x)&\sim&
\int_{q-\Lambda''}^{q+\Lambda''}\frac{dp}{2\pi}\ e^{i(p-q)x}R_s(p)\ ,\
\Lambda''\ll |q|.
\eea
Substituting the decomposition \fr{decomp} into our expression of the
Hamiltonian density \fr{HLE} and dropping oscillatory contributions
under the integral, we arrive at the following mobile impurity model 
\bea 
H^{(0)}_{\rm MIM}&=& \int dx\sum_{\alpha=c,s}-iv_\alpha'\left(
r^\dagger_\alpha \partial_x r^{\phantom\dagger}_\alpha
-l^\dagger_\alpha \partial_xl^{\phantom\dagger}_\alpha\right)\nn
&+&\int dx\ \chi^\dagger_s\left(\varepsilon_s-iu_s\partial_x+\dots\right)\chi_s\ .
\label{HLE_MIM}
\eea
Here we have introduced notations
$\varepsilon_s=\eps_s(k_F-q)\approx v_s'q+\zeta q^3+\dots$ and
$u_s=-\left.\frac{d\eps_s}{dk}\right|_{k_F-q}\approx v_s^\prime +
3\zeta q^2$. 

Next we need to work out the projection of the operator ${\cal O}_s$ 
on low-energy ($r_s$) and impurity ($\chi_s$) degrees of freedom.
Using that $r_s(x)$, $l_s(x)$ and $\chi_s(x)$ are slowly varying
fields, we can approximate the string operator as 
\bea
{\cal
  O}_s(x)&\sim&\left[r_s(x)+e^{-iqx}\chi^\dagger_s(x)\right]
e^{\frac{\pi}{2}\left[\frac{e^{-iqx}}{q}r^\dagger_s(x)\chi^\dagger_s(x)
-{\rm h.c.}\right]}\nn
&\times& e^{-\frac{i\pi}{2}\int_{-\infty}^x dx'\left[r^\dagger_s(x)r_s(x)
+l^\dagger_s(x)l_s(x)-\chi^\dagger_s(x)\chi_s(x)\right]}.\nn
\label{highlow}
\eea
Here the second term is the contribution of the string arising from the
upper boundary of integration $x$. 
By virtue of the presence of the strongly oscillatory factor $e^{iqx}$
in the expression \fr{holepart} for the spectral function, the leading
contribution to the spectral function arises from the part of ${\cal
  O}_s(x)$ proportional to $e^{-iqx}$ 
\be
{\cal O}_s(x)={\cal O}_s^{(q)}(x)e^{-iqx}+\dots
\ee
In terms of this component we have
\bea
A_<(\omega,k_F-q)&\sim&-i\int_0^\infty\! dt\int_{-\infty}^\infty\!\!\! dx\
e^{i\omega t}\langle{\cal O}_c^\dagger(0,0){\cal O}_c(t,x)\rangle\nn
&&\qquad\qquad\times
\langle{{\cal O}_s^{(q)}}^\dagger(0,0){\cal O}_s(t,x)\rangle\ .
\label{holepart2}
\eea 
In order to isolate the desired contribution, we expand the second factor 
in \fr{highlow} 
\begin{widetext}
\bea
{\cal O}_s^{(q)}(x)&\sim& e^{-iqx}\bigg\lbrace r_s(x)\Big\lbrack\frac{\pi}{2q}
r_s^\dagger(x)\chi^\dagger_s(x)+\ldots\Big\rbrack
+\chi_s^\dagger(x)\Big\lbrack1-\frac{\pi^2}{8q^2}
\lbrace\chi_s(x)r_s(x),r_s^\dagger(x)\chi^\dagger_s(x)\rbrace
+\ldots\Big\rbrack\bigg\rbrace\nn
&\times& e^{-\frac{i\pi}{2}\int_{-\infty}^x dx'\left[r^\dagger_s(x^\prime)r_s(x^\prime)
+l^\dagger_s(x^\prime)l_s(x^\prime)-\chi^\dagger_s(x^\prime)\chi_s(x^\prime)\right]}.
\label{highlow2}
\eea
\end{widetext}

In order to proceed further, it is convenient to bosonize
the low-energy degrees of freedom associated with $r_s$,
$l_s$ using \fr{boso_sc}:
\bea
{\cal
  O}_s^{(q)}(x)&\sim&e^{-iqx}\chi^\dagger_s(x)\left[b_0+b_1\partial_x\varphi_s^*(x)+\ldots\right]\nn
&\times& 
 e^{\frac{i}{4\sqrt{2}}\Phi_s^*(x)
+\frac{i\pi}{2}\int_{-\infty}^x dx'\,\chi^\dagger_s(x^\prime)\chi_s(x^\prime)}.
\eea
This can be simplified further using the appropriate operator product
expansions. In order to determine the threshold singularity it is
sufficient to retain only the term with the lowest scaling dimension
at low energies, which is
\be
{\cal O}_s^{(q)}(t,x)\sim e^{-iqx}
\chi^\dagger_s(t,x)e^{\frac{i}{4\sqrt{2}}\Phi_s^*(t,x)}.
\label{Os}
\ee
The two-point correlator of this operator is
\bea
&&\langle{{\cal O}^{(q)}_s}^\dagger(0,0){\cal O}^{(q)}_s(t,x)\rangle\nn
&&\sim e^{-iqx}
\ (x^2-{v_s'}^2t^2)^{-\frac{1}{16}}\
e^{i\varepsilon_st}\frac{\sin\big[\Lambda''(x-u_st)\big]}{
\pi(x-u_st)}.
\label{OsOs}
\eea
Sending the cutoff $\Lambda''$ to infinity turns the last term  into a
delta function $\delta (x-u_st)$. Substituting the resulting expression
for \fr{OsOs} and the charge sector contribution \fr{OcOc} into the
expression \fr{holepart} for the hole spectral function and then
carrying out the space and time integrals, we arrive at the following
result for the threshold behaviour
\be
A_<(\omega,k_F-q)\propto |\omega+\varepsilon_s|^{-1/4}.
\ee
As expected there is a threshold singularity. The exponent is seen to
be momentum independent. As we will see, this is particular to the LE point.
\section{Mobile impurity model away from the LE point\label{sec:MIMaway}}
We now wish to generalize the above analysis to the Luttinger
liquid phase surrounding the LE point. We will assume that 
\begin{enumerate}
\item{} the spinon mass term is fine-tuned to zero, i.e. $g_{s,1}=0$;
\item{} the four fermion interactions in the spin and charge sectors
are \emph{attractive}, i.e. $g_{c,0},g_{s,0}<0$, and sizeable.
\end{enumerate}
Under these assumptions holons and spinons remain gapless, and the
$g_1$ term in \fr{perturbations} is irrelevant so that we can drop it
at low energies. Focussing again on the single-electron spectral
function, using the decomposition \fr{decomp}, and finally bosonizing
the low-energy spin and charge degrees of freedom, we arrive at a
mobile impurity model of the form
\be
H_{\rm MIM}=\int dx\left[
\sum_{\alpha=c,s}{\cal H}_\alpha+{\cal H}_{\rm imp}+{\cal
  H}_{\rm int}\right],
\ee
\bea
{\cal H}_\alpha&=&\frac{v_\alpha}{16\pi}
\left[\frac{1}{2K_\alpha}\big(\partial_x\Phi_\alpha^*\big)^2
+2K_\alpha\big(\partial_x\Theta_\alpha^*\big)^2
\right] ,\nn
{\cal H}_{\rm  imp}&=&\chi^\dagger_s
\left( \varepsilon_s-i {u}_s\partial_x\right)\chi_s\ ,\nn 
{\cal H}_{\rm int}&=&\chi^\dagger_s\chi_s\left[\sum_\alpha
f_\alpha(q)\partial_x\varphi^*_\alpha+
\bar{f}_\alpha(q)\partial_x\bar\varphi^*_\alpha\right].
\label{MIM_away}
\eea
Here we have dropped all terms that do not affect the threshold
exponent  and retained the same parameterization of the
impurity part of the Hamiltonian, although the actual values of
$\epsilon_s$ and $u_s$ are of course not the same as the LE point.
The Luttinger parameter in the spin sector varies from
$K_s=1/2$ at the LE point to $K_s=1$ in the SU(2)-invariant
limit. The charge Luttinger parameter equals $K_c=1/2$ at the LE
point, and varies with doping and interaction strength otherwise.
We note that   close to the LE point (in the sense that $g_{0,c}$,
$g_{0,s}$ are small), there is an additional contribution to ${\cal
  H}_{\rm int}$ of the form
$\chi^\dagger_s\chi_s\cos(\Phi_s^*/\sqrt{2})$. The analysis of this
case is very interesting (see e.g. Ref.~\onlinecite{ludwig} for a related
problem), but beyond the scope of our work. The functions $f_\alpha(q)$,
$\bar{f}_\alpha(q)$ as well as the parameters $v_\alpha$, $K_\alpha$,
 $\varepsilon_s$, $u_s$ depend on the microscopic
details of the particular lattice realization of our field theory. 
We will show below how they can be fixed either numerically in the
generic case or analytically when our theory is applied to the Hubbard
model.

The mobile impurity model \fr{MIM_away} can now be analyzed by
standard methods \cite{Schotte,SIG3}. The interaction between the
impurity and the low-energy degrees of freedom can be removed through
a unitary transformation 
\be
U=e^{-i\int_{-\infty}^\infty dx
\sum_{\alpha}\left[\gamma_\alpha\varphi^*_\alpha(x)
+\bar\gamma_\alpha\bar\varphi^*_\alpha(x)\right]\chi_s^\dagger(x)\chi_s(x)}.
\ee
The transformed spin impurity field equals
\bea
d_s(x)&=&U\chi_s(x)U^\dagger\nonumber\\
&=&
\chi_s(x)e^{i\sum_\alpha [\gamma_\alpha\varphi^*_\alpha(x)+
\bar\gamma_\alpha\bar\varphi^*_\alpha(x)]}\nn
&&\times 
e^{-i\pi\sum_\alpha(\gamma_\alpha^2-{\bar\gamma}_\alpha^2) C(x)},\
\eea
while the chiral spin and charge Bose fields transform as
\bea
\varphi^\circ_\alpha(x)=U\varphi_\alpha^*(x)U^\dagger=
\varphi_\alpha^*(x)-2\pi\gamma_\alpha C(x),\nn
\bar\varphi^\circ_\alpha(x)=U\bar\varphi_\alpha^*(x)U^\dagger=
\bar\varphi_\alpha^*(x)+2\pi\bar\gamma_\alpha C(x),
\eea
where 
\bea
C(x)&=&
\int_{-\infty}^{\infty} dy\ {\rm sgn}(x-y)\chi^\dagger_s(y)\chi_s^{\phantom\dagger}(y)\ .
\eea
We note that
\bea
\partial_x\varphi^\circ_\alpha(x)&=&\partial_x
\varphi_\alpha^*(x)-4\pi\gamma_\alpha \chi^\dagger_s(x)\chi^{\phantom\dagger}_s(x),\nn
\partial_x\bar\varphi^\circ_\alpha(x)&=&\partial_x
\bar\varphi_\alpha^*(x)+4\pi\bar\gamma_\alpha \chi^\dagger_s(x)\chi^{\phantom\dagger}_s(x).
\label{twist}
\eea
Adjusting the parameters $\gamma_\alpha$, $\bar\gamma_\alpha$ such
that
\bea
\begin{pmatrix}
f_\alpha\\
\bar f_\alpha
\end{pmatrix}&=&
\begin{pmatrix}
u-v_{\alpha}^+ &-v_{\alpha}^-\\
v_{\alpha}^- & v_{\alpha}^++u
\end{pmatrix}
\begin{pmatrix}
\gamma_\alpha\\
\bar \gamma_\alpha
\end{pmatrix},\eea
with
\bea
v_\alpha^\pm&=&\frac{v_\alpha}{2}\left(2K_\alpha\pm\frac{1}{2K_\alpha}\right),
\eea
the impurity decouples in the new variables
\bea
H_{\rm MIM}&=&\int dx\left[
\sum_{\alpha=c,s}{\cal H}'_\alpha+{\cal H}'_{\rm imp}\right],\nn
{\cal H}'_\alpha&=&\frac{v_\alpha}{16\pi}
\left[\frac{1}{2K_\alpha}\big(\partial_x\Phi^\circ_\alpha\big)^2
+2K_\alpha\big(\partial_x\Theta^\circ_\alpha\big)^2
\right] ,\nn
{\cal H}'_{\rm  imp}&=&d^\dagger_s
\left(\tilde\varepsilon_s-iu_s\partial_x\right)d^{\phantom\dagger}_s\ .
\label{Htrans}
\eea 

The interaction between the mobile impurity and the Luttinger liquid
degrees of freedom is now encoded in the boundary conditions of the
transformed Bose fields $\Phi^\circ_\alpha$, $\Theta^\circ_\alpha$,
which are ``twisted'' by the presence of the impurity, see
e.g. \fr{twist}. The negative-frequency part of the single-electron
spectral function is again given by \fr{holepart2}, where
\bea
{\cal O}_c(x){\cal O}_s^{(q)}(x)\sim
e^{-iqx}\chi^\dagger_se^{\frac{i}{4\sqrt{2}}\Phi_s^*}
e^{-\frac{i}{\sqrt{2}}\varphi_c^*}e^{\frac{i}{4\sqrt{2}}\Phi_c^*}\ .
\eea
In terms of the transformed fields this reads
\bea
{\cal O}_c(x){\cal O}_s^{(q)}(x)&\sim&
e^{-iqx}d^\dagger_s\ e^{i(\gamma_s+\frac{1}{4\sqrt{2}})\varphi_s^\circ
+i(\bar\gamma_s+\frac{1}{4\sqrt{2}})\bar\varphi_s^\circ}\nn
&&\times\quad e^{i(\gamma_c-\frac{3}{4\sqrt{2}})\varphi_c^\circ
+i(\bar\gamma_c+\frac{1}{4\sqrt{2}})\bar\varphi_c^\circ}\ .
\eea
The threshold behaviour of the hole spectral function can now be
calculated in the same way as at the LE point. The result is
\bea
A_<(\omega,k_F-q)&\sim& \frac{1}{|\omega+\eps_0|^\mu}\ ,
\eea
where the exponent $\mu$ is given by
\bea
\mu&=&1-2(\nu_{c,+}^2+\nu_{c,-}^2+\nu_{s,+}^2+\nu_{s,-}^2),\nn
\nu_{c,\pm}&=&\sqrt{\frac{K_c}{2}}\Big[\gamma_c+\bar\gamma_c-\frac{1}{2\sqrt{2}}\Big]
\pm\frac{1}{\sqrt{8K_c}}\Big[\gamma_c-\bar\gamma_c-\frac{1}{\sqrt{2}}\Big],\nn
\nu_{s,\pm}&=&\sqrt{\frac{K_s}{2}}\Big[\gamma_s+\bar\gamma_s+\frac{1}{2\sqrt{2}}\Big]
\pm\frac{1}{\sqrt{8K_s}}\big[\gamma_s-\bar\gamma_s\big].
\label{nus}
\eea
As $\gamma_\alpha$, $\bar\gamma_\alpha$ are functions of $q$, the
threshold exponent is now generally momentum dependent. However, it is shown in
Appendix \ref{app:Su2} that spin rotational SU(2) symmetry in the limit $K_s\to
1$ enforces the particular values
\be
\gamma_s=\bar\gamma_s=-\frac{1}{4\sqrt{2}},\label{fixgammasSU2}
\ee
for any value of $q$.

\subsection{Relation of \texorpdfstring{$\gamma_\alpha$,
      $\bar\gamma_\alpha$}{LG} to finite-size energy spectra}
An obvious question is whether there is a way of directly determining
the parameters $\gamma_\alpha$, $\bar\gamma_\alpha$ for a given
microscopic lattice model. To that end, let us consider the spectrum
of our mobile impurity model on a large, finite ring of circumference $L$.
The mode expansions of the Bose fields $\varphi_\alpha^*$ ,
$\bar\varphi_\alpha^*$ are
\begin{widetext}
\bea
\varphi^*_\alpha(x)&=&\varphi_{\alpha,0}^*+\frac{x}{L}Q^*_\alpha+\sum_{n=1}^\infty
\sqrt{\frac{2}{n}}
\left[e^{i\frac{2\pi n}{L}x}a_{\alpha,R,n}+e^{-i\frac{2\pi
    n}{L}x}a^\dagger_{\alpha,R,n}\right] ,\nn
\bar{\varphi}^*_\alpha(x)&=&\bar{\varphi}^*_{\alpha,0}+\frac{x}{L}
\bar{Q}^*_\alpha+\sum_{n=1}^\infty\sqrt{\frac{2}{n}}\left[  
e^{-i\frac{2\pi n}{L}x}a_{\alpha,L,n}+e^{i\frac{2\pi n}{L}x}a^\dagger_{\alpha,L,n}\right] .
\label{modeexp}
\eea
\end{widetext}
Here $Q^*_\alpha$, $\bar{Q}^*_\alpha$, $\varphi^*_{\alpha,0}$ and
$\bar\varphi^*_{\alpha,0}$ are zero mode operators with commutation
relations
\be
[\varphi^*_{\alpha,0},Q^*_\alpha]=-4\pi i=
-[\bar{\varphi}^*_{\alpha,0},\bar{Q}^*_\alpha].
\ee
The eigenvalues $q_\alpha$, $\bar q_\alpha$ of the zero mode operators
$Q^*_\alpha$, $\bar Q^*_\alpha$ depend on the boundary conditions on the
fields $\varphi^*_{\alpha}$, $\bar\varphi^*_{\alpha}$, which on
general grounds will depend on whether or not a mobile impurity is present.
In presence of the impurity the finite-size spectrum of
\fr{MIM_away} has the following structure 
\be
E=E_{\rm GS}+E_{\rm imp}+\Delta E_{\rm LL}+o(L^{-1}).
\label{EMIM}
\ee
Here $E_{\rm imp}$ is the contribution of the impurity to the
energy. On general grounds it will have the following expansion in
terms of system size
\be
E_{\rm imp}=E_{\rm imp}^{(0)}+\frac{1}{L}E_{\rm imp}^{(1)}+o(L^{-1}).
\ee
The other contribution to \fr{EMIM} arises from the Luttinger-liquid
part of the theory. Applying the mode expansions to the transformed
Hamiltonian \fr{Htrans} we obtain
\begin{widetext}
\bea
\Delta E_{\rm LL}&=&\sum_{\alpha=c,s}
\frac{2\pi v_\alpha}{L}\left[\frac{1}{4K_\alpha}\left(\frac{q_\alpha+\bar
    q_\alpha}{4\pi}-\gamma_\alpha+\bar\gamma_\alpha\right)^2
+K_\alpha\left(\frac{q_\alpha- \bar
  q_\alpha}{4\pi}-\gamma_\alpha-\bar\gamma_\alpha\right)^2
+\sum_{n>0}n\left[M_{n,\alpha}^++M_{n,\alpha}^-\right]
\right].
\label{EMIM2}
\eea
\end{widetext}
Here $q_\alpha$, $\bar q_\alpha$ and $M_{n,\alpha}^\pm$ are ``quantum
numbers'' characterizing a particular low-energy excitation. Their
quantization conditions depend on the boundary conditions for the spin
and charge Bose fields \fr{modeexp} in presence of a high-energy
mobile impurity. These can be worked out by considering the
``minimal'' excitation that can be made in the sector where the mobile
impurity is present. This sector is reached by acting with the operator
\be
{\cal O}_c(x){\cal O}_s^{(q)}(x)
=e^{-iqx}\chi^\dagger_se^{\frac{i}{4\sqrt{2}}\Phi_s^\circ}
e^{-\frac{i}{\sqrt{2}}\varphi_c^\circ}e^{\frac{i}{4\sqrt{2}}\Phi_c^\circ}
\label{OcOs2}
\ee
on the ground state $|\psi_0\rangle$. The latter is characterized by being
annihilated by $a_{\alpha,R,n}$, $a_{\alpha,L,n}$, $Q^*_\alpha$ and
$\bar Q^*_\alpha$. The quantum numbers of the ``minimal'' excitation are
found by noting that
\bea
Q^*_\alpha{\cal O}_c(x){\cal O}_s^{(q)}(x)|\psi_0\rangle&=&
[Q^*_\alpha,{\cal O}_c(x){\cal O}_s^{(q)}(x)]|\psi_0\rangle\nn
&=&q^{(0)}_\alpha{\cal O}_c(x){\cal O}_s^{(q)}(x)|\psi_0\rangle,
\eea
where
\bea
q^{(0)}_c=\frac{3\pi}{\sqrt{2}},\ \bar q^{(0)}_c=\frac{\pi}{\sqrt{2}},\
q^{(0)}_s=-\bar q^{(0)}_s=-\frac{\pi}{\sqrt{2}}.
\label{q0}
\eea
Let us denote the lowest energy state with these quantum numbers by
\be
|q_c^{(0)},\bar q_c^{(0)},q_s^{(0)},\bar q_s^{(0)}\rangle.
\label{impvac}
\ee
Higher excited states in the ``impurity sector'' can be obtained for
example by making particle-hole excitations on top of the state
\fr{impvac}. The energies of such states are given by \fr{EMIM2} by
choosing \fr{q0} and in addition taking some of the $M^\pm_{n,\alpha}$
different from zero. Crucially, the values of $\gamma_\alpha$,
$\bar\gamma_\alpha$ are the same as for the state \fr{impvac}.
For practical purposes, states with the same momentum as \fr{impvac}
but different spin and charge quantum numbers might be of particular
interest as they are lowest energy states in certain sectors of
quantum numbers and can therefore be more easily targeted in DMRG computations. 
The quantum numbers for such states are
\bea
q_\alpha&=&q^{(0)}_\alpha+\frac{\pi}{\sqrt{2}}(3m_\alpha+\bar
m_\alpha),\nn
\bar q_\alpha&=&\bar q^{(0)}_\alpha+\frac{\pi}{\sqrt{2}}(3\bar
m_\alpha+\ m_\alpha),
\label{Qeigen}
\eea
where $m_\alpha$, $\bar m_\alpha$ are integers. This follows from the
mode expansion and the requirement, that the bosonized expressions for
$R_\up(x)$, $R_\down(x)$, $L_\up(x)$, $L_\down(x)$ must be single valued.
Using that at low energies the charge and spin densities and currents
are given by 
\bea
\rho_\alpha(x)&=&-\frac{1}{\pi\sqrt{8}}\partial_x\Phi_\alpha^\circ(x)\ ,\nn
j_\alpha(x)&=&-\frac{1}{\pi\sqrt{8}}\partial_x\Theta_\alpha^\circ(x)\ ,
\eea
we can identify
\begin{itemize}
\item{} $(m_c+\bar m_c)/2$ is the difference in charge (particle number)
between the excitation and the state \fr{impvac};
\item{} $(m_c-\bar m_c)/2$ is the number of charge fermions
transferred from the right-moving to the left-moving branch;
\item{} $(m_s+\bar m_s)/2$ is the difference in the number of down spins
between the excitation and the state \fr{impvac};
\item{} $(m_s-\bar m_s)/2$ is the number of spin fermions
transferred from the right-moving to the left-moving branch.
\end{itemize}

The values of $\gamma_\alpha$, $\bar\gamma_\alpha$ can now in
principle be extracted by numerically computing finite-size energy
levels by a method such as momentum-space DMRG \cite{xiang,schollwock}.
The procedure is outlined in the following.

\noindent
(1) The velocities $v_\alpha$ and Luttinger parameters $K_\alpha$
are bulk properties and can be determined by standard methods. We will
assume them to be known quantities in the following. We further denote
the number of
particles and down spins in the ground state by $N_{\rm GS}$ and
$M_{\rm GS}$ respectively.

\noindent
(2) We then numerically compute the lowest excitation with
momentum $q$  and quantum numbers $N=N_{\rm GS}-1$, $M=M_{\rm
  GS}-1$. Its energy is  
\bea
E_{-1,-1}&=&E_{\rm GS}+E^{(0)}_{\rm imp}+\frac{1}{L}E^{(1)}_{\rm imp}\nn
&+&\Delta E_{\rm LL}(q_c^{(0)},\bar q_c^{(0)},
q_s^{(0)},\bar q_s^{(0)})+o(L^{-1}).
\eea

\noindent
(3) Next we compute the lowest excitations with momentum $q$ but
different values of $N$ and $M$. For example, choosing
\be
N=N_{\rm GS}-1\ ,\quad M=M_{\rm GS}-2,
\ee
gives the excited state characterized by
\be
m_c=-\bar m_c=1\ ,\quad
m_s=-2\ ,\bar m_s=0.
\label{hsss}
\ee
The zero-mode eigenvalues follow from \fr{Qeigen}
\bea 
q_c^{(1)}&=&\frac{5\pi}{\sqrt{2}}\ ,\
\bar q_c^{(1)}=-\frac{\pi}{\sqrt{2}}\ ,\nn
q_s^{(1)}&=&-\frac{7\pi}{\sqrt{2}}\ ,\
\bar q_s^{(1)}=-\frac{\pi}{\sqrt{2}}.
\eea
The corresponding energy is
\bea
E_{-1,-2}&=&E_{\rm GS}+E^{(0)}_{\rm imp}+\frac{1}{L}E^{(1)}_{\rm imp}\nn
&&+\Delta E_{\rm LL}(q_c^{(1)},\bar q_c^{(1)},
q_s^{(1)},\bar q_s^{(1)})+o(L^{-1}).
\eea
Here we have asserted that the change in the impurity contribution to
the energy is of higher order in $L^{-1}$. This has been shown for the
case of the Hubbard model using methods of integrability in
Ref. \onlinecite{FHLE}. We believe that this continues to hold true in
general, because $E^{(1)}_{\rm imp}$ is sensitive only to the values
of the parameters $\gamma_\alpha$, $\bar\gamma_\alpha$, which are the
same for all excitations we are considering. 

The point is that by considering energy differences like
\bea
E_{-1,-2}-E_{-1,-1}&=&\Delta E_{\rm LL}(q_c^{(1)},\bar q_c^{(1)},
q_s^{(1)},\bar q_s^{(1)})\nn
&-&\Delta E_{\rm LL}(q_c^{(0)},\bar q_c^{(0)},
q_s^{(0)},\bar q_s^{(0)})+o(L^{-1})\nn
\eea
we obtain a set of equations in which the only unknown parameters are
the $\gamma_\alpha$, $\bar\gamma_\alpha$. This provides a numerical
method for determining them in a general lattice model. For integrable
theories like the Hubbard model, analytical techniques are
available and we discuss this case next.

\subsection{Hubbard model\label{sec:Hubbard}}
The finite-size spectrum in presence of a high-energy spinon
excitation was calculated for the case of the Hubbard model using the
Bethe Ansatz solution in Ref.~\onlinecite{FHLE}. The result for the lowest
excited state above the spinon threshold is
\begin{widetext}
\bea
E&=&E_{\rm GS}-\eps_s(\Lambda^h)-\frac{1}{L}\eps'_s(\Lambda^h)\delta\Lambda^h\nn
&+&\frac{2\pi  v_c}{L}
\left[\frac{(\Delta N_c-N_c^{\rm imp})^2}{8K_c}
+2K_c\left({D}_c-D^{\rm imp}_c+\frac{{D}_s}{2}\right)^2\right]
+\frac{2\pi
  v_s}{L}
\left[\frac{\left(\Delta {N}_s-\frac{1}{2}\Delta {N}_c-\frac{1}{2}\right)^2}{2}
+\frac{D_s^2}{2}
\right]+o(L^{-1}),
\label{EBA}
\eea
\end{widetext}
where
\be
D_s=D_c=0\ ,\quad \Delta N_c=-1\ ,\quad \Delta N_s=0.
\label{DN}
\ee
The contribution
$-\eps_s(\Lambda^h)-\frac{1}{L}\eps'_s(\Lambda^h)\delta\Lambda^h$ is
the finite-size energy of the impurity. The velocities $v_\alpha$,
$K_\alpha$ as well as the quantities $N_c^{\rm imp}$ and $D_c^{\rm
imp}$ are expressed in terms of solutions to coupled linear 
integral equations, and in practice are easily calculated
numerically with very high precision.
By construction the quantum numbers \fr{DN} correspond to our minimal
excited state \fr{impvac}. By matching the Luttinger liquid part of
the energy to \fr{EMIM} we then obtain the following results for the
parameters $\gamma_\alpha$, $\bar\gamma_\alpha$
\bea
\gamma_c+\bar\gamma_c&=&\frac{1}{2\sqrt{2}}-\sqrt{2}D_c^{\rm imp},\
\gamma_c-\bar\gamma_c=-\frac{N_c^{\rm imp}}{\sqrt{2}},\nn
\gamma_s&=&\bar\gamma_s=-\frac{1}{4\sqrt{2}}.
\label{qgamma}
\eea
One subtlety to keep in mind when making contact between the Bethe
Ansatz calculation and \fr{EMIM} is that the former refers only to
highest weight states of the SU(2)$\otimes$ SU(2) symmetry algebra of
the Hubbard model\cite{EKS1,EKS2}. Descendant states need to be taken
into account separately. Substituting \fr{qgamma} in the expressions
for the quantities $\nu_\alpha^\pm$ \fr{nus}, we find
\be
\nu_s^\pm=0,\
\nu_c^\pm=-\sqrt{K_c}D_c^{\rm imp}\mp\frac{1+N_c^{\rm imp}}{4\sqrt{K_c}}.
\ee
Finally, the threshold exponent is obtained from \fr{nus}
\be
\mu=1-\frac{(1+N_c^{\rm imp})^2}{4K_c}-4K_c(D_c^{\rm imp})^2.
\ee
This agrees with what was found in Ref.~\onlinecite{FHLE} using the
approach of Schmidt, Imambekov and Glazman \cite{SIG1,SIG2}, as well
as with the exponents reported previously in
Refs.~\onlinecite{carmelo1,carmelo2,carmelo3}.

\subsubsection{Other excited states}
The Bethe Ansatz result \fr{EBA} can be applied to other excited
states as well. In particular, the holon plus three spinon excitation
considered in Ref.~\onlinecite{FHLE} gives rise to an excitation with the
same momentum and quantum numbers
\be
\Delta N_c=-1=\Delta N_s\ ,\ D_s=-1\ ,\ D_c=\frac{1}{2}.
\ee
The corresponding state in our mobile impurity model has quantum
numbers \fr{hsss}, and the energies calculated from \fr{EMIM} and
\fr{EBA} agree as they must.

\section{Relation to the approach of Schmidt, Imambekov and Glazman}

The method of Refs. \onlinecite{SIG1,SIG2} is based on a different
prescription for defining fermionic quasiparticles in the charge and
spin sectors. We now summarize the main steps of this approach and
compare them to our framework. The starting point is the standard
bosonized description (\ref{LLwithchargeandspin}) of the spinful
fermion model under consideration. One then introduces the chiral components
$\phi_\alpha,\bar \phi_\alpha$ by  
\bea 
\Phi_\alpha(x)&=&\sqrt{K_\alpha}(\phi_\alpha+\bar\phi_\alpha),\\
\Theta_\alpha(x)&=&\frac1{\sqrt{K_\alpha}}(\phi_\alpha-\bar\phi_\alpha).
\eea
These chiral bosons diagonalize the spinful Luttinger model
(\ref{LLwithchargeandspin}) in the form
\be 
\mc{H}_{\text{LL}}(x)=\sum_{\alpha=c,s}\frac{v_\alpha}{8\pi}\left[(\partial_x\phi_\alpha)^2+(\partial_x\bar \phi_\alpha)^2\right].
\ee
Next, in analogy with the spinless case in  Eqs. (\ref{quasiR}) and (\ref{quasiL}), one defines the quasiparticle operators
\bea
\tilde R_\alpha(x)\sim e^{-\frac{i}{\sqrt2}\phi_\alpha(x)}, \quad \tilde L_\alpha(x)\sim e^{\frac{i}{\sqrt2}\bar\phi_\alpha(x)}.\label{quasiRLcs}
\eea
As in the spinless case, this prescription removes the marginal
interactions between quasiparticles, rendering them asymptotically
free in the low-energy limit. However, these quasiparticles cannot be
identified with the finite energy elementary excitations in integrable
models, because they carry fractional quantum numbers:
\be 
[q_\alpha,\tilde R_\alpha^\dagger(x)]=\sqrt{2K_\alpha} \tilde R^\dagger_\alpha,\quad [q_\alpha,\tilde L_\alpha^\dagger(x)]=\sqrt{2K_\alpha} \tilde L^\dagger_\alpha.
\ee
The only exception is the LE point $K_\alpha=1/2$, where the operators
$\tilde R_\alpha,\tilde L_\alpha$ in fact carry the same quantum numbers as
our spin and charge fermions.

Despite the lack of correspondence with long-lived excitations in
(nearly) integrable models, the approach of Refs.
\onlinecite{SIG1,SIG2} allows one to compute threshold exponents, as 
long as the parameters in the effective impurity model are adjusted
appropriately. The situation is analogous to the two ways of
determining threshold exponents discussed above in the spinless
fermion case. 

In order to facilitate a direct comparison with our
approach, we briefly review how to express the electron operator in
order to calculate the threshold exponent in the single-particle
spectral function within the approach of Refs.~\onlinecite{SIG1,SIG2}. We assume again
that the lower threshold of the support  corresponds to  an excitation
with a single high-energy spinon. First, the right-moving spin quasiparticle in
Eq. (\ref{quasiRLcs}) is  projected into low-energy and impurity
subbands:\be 
\tilde R_s(x)\sim \tilde r_s(x)+e^{-iqx}\tilde \chi_s^\dagger(x).
\ee
One then rewrites the electron operator in terms of the quasiparticles
defined in (\ref{quasiRLcs}). As this differs from ours, cf
Eq. (\ref{spinlessRalpha}), the string-operator part in the expression
for the electron operator is also different
\bea
R_\uparrow(x)&\sim& e^{-iqx}\tilde \chi_s^\dagger(x)
e^{\frac{i}{\sqrt2}\phi_s}\prod_{\alpha=c,s}
e^{-\frac{i}{4}\left(\sqrt{K_\alpha}+\frac1{\sqrt{K_\alpha}}\right)\phi_\alpha}\nn
&&\qquad\qquad\qquad\times\ e^{-\frac{i}{4}\left(\sqrt{K_\alpha}-\frac1{\sqrt{K_\alpha}}\right)\bar\phi_\alpha}.
\eea
The next step is to write down an effective impurity model, analogous to Eq. (\ref{MIM_away}), but using $\tilde \chi_s$ as the impurity. After performing a unitary transformation that removes the coupling between $\tilde \chi_s$ and the low-energy modes, the electron operator becomes\bea
R_\uparrow(x)&\sim& e^{-iqx}\tilde d_s^\dagger(x)   e^{-\frac{i}4\left(\sqrt{K_s}+\frac1{\sqrt{K_s}}-2\sqrt2-4\gamma_s^\prime\right)\phi_s}\nonumber\\
&&\times e^{-\frac{i}{4}\left[\left(\sqrt{K_s}-\frac1{\sqrt{K_s}}-4\bar\gamma_s^\prime\right)\bar\phi_s+\left(\sqrt{K_c}+\frac{1}{\sqrt{K_c}}-4\gamma_c^\prime\right)\phi_c\right]}\nonumber\\
&&\times e^{-\frac{i}{4}\left[\left(\sqrt{K_c}-\frac1{\sqrt{K_c}}-4\bar\gamma_c^\prime\right)\bar\phi_c\right]},\label{electronIG}
\eea
where $\tilde d_s=U\tilde\chi_sU^\dagger$ is the free impurity field
for the quasiparticle with fractional charge, and
$\gamma^\prime_\alpha,\bar\gamma_\alpha^\prime$ are the parameters of
the unitary transformation, which are not the same as
$\gamma_\alpha,\bar\gamma_\alpha$ discussed in Section
\ref{sec:MIMaway}.  

At the LE point, we set $K_\alpha=1/2$ and
$\gamma^\prime_\alpha=\bar\gamma_\alpha^\prime=0$ and
Eq. (\ref{electronIG}) reduces to 
\be 
R_\uparrow(x)\sim e^{-iqx}\tilde d_s^\dagger(x)
e^{-\frac{i}{\sqrt{2}}\phi_c}e^{\frac{i}{4\sqrt{2}}(\phi_s+\bar\phi_s+\phi_c+\bar\phi_c)}. 
\ee
This result agrees with the refermionization in Section
\ref{sec:LEexponent}, since at the LE point $\tilde
d_\alpha=d_\alpha=\chi_\alpha$, $\phi_\alpha=\varphi_\alpha^*$,   and
$\bar \phi_\alpha=\bar\varphi_\alpha^*$; thus, the two approaches
coincide.  

Moving away from the LE point, the expressions for physical operators
in terms of impurity and low-energy fields will  in general  be
different in the two approaches. However, the results for the edge
exponents are still consistent because the difference in string
operators can be accommodated by the parameters of the unitary
transformation and by imposing  proper boundary conditions on the
bosonic fields. For instance, imposing SU(2) symmetry in the
approach of Refs. \onlinecite{SIG1,SIG2} leads to the requirements
\be
\gamma_s^\prime=\sqrt2-1,\quad \bar \gamma_s^\prime=0.
\ee
This should be contrasted with Eq. (\ref{fixgammasSU2}).


\section{Realizing the LE point}
\label{sec:LEP}
We have seen that a good starting point for understanding threshold
singularities in dynamical response functions is the LE point for both
charge and spin. In section~\ref{sec:LEP_FT} we considered properties
of the LE point in the field theory limit. An obvious question raised
by these considerations is whether it is possible to realize the LE
point in practice in a lattice model of interacting spinful
fermions. We now investigate this issue in some detail and present a
number of preliminary results.

\subsection{Lattice model}
\label{sec:latticemodel}
As discussed in section \ref{ssec:stability}, realizing the LE point
at sufficiently low energies in an extended Hubbard model of the kind
\fr{HEHubb} requires the fine-tuning of (at least) four
parameters:
\be
K_c = K_s = \frac{1}{2}\ ,\quad g_{s,1} = g_1 = 0.
\label{finetunings}
\ee
For the Hubbard model in zero magnetic field, spin rotational symmetry
fixes $K_s=1$, while $K_c$ varies with both band filling and
interaction strength $U$. In particular, it is well known
\cite{woynarovich,frahmkorepin} that $K_c=\frac{1}{2}$ is obtained in
the $U\to\infty$ limit of the Hubbard model
\cite{ogatashiba,penc}. Values $K_s<1$ can be realized by adding
spin-dependent interactions that break the spin SU(2) symmetry, but
retain spin inversion symmetry. The latter is crucial for avoiding
marginal interactions between spin and charge sectors that lead to a
more complicated conformal spectrum involving a dressed charge matrix
\cite{hubbardbook,woynarovich,frahmkorepin}. 

A minimal lattice model that may allow us to fulfil the conditions in
(\ref{finetunings}) is
\begin{eqnarray}
H&=&-t\sum_{j=1}^{L-1}\sum_{\sigma}(c_{j,\sigma}^\dagger 
c_{j+1,\sigma}^{\phantom\dagger}+{\rm h.c.})+U
\sum_{j=1}^Ln_{j,\uparrow}n_{j,\downarrow}\nn 
&&+\sum_{r=1}^2 V_r \sum_{j=1}^{L-r} n_{j} n_{j+r}+J_1^z
\sum_{j=1}^{L-1} S_j^z S_{j+1}^z. 
\label{Hext}
\end{eqnarray}
In anticipation of the DMRG computations of energy levels reported
below we have imposed open boundary conditions. In the following we
set $t=1$, i.e. measure all energies in units of the hopping parameter.
The idea is then to try to adjust the four interaction strengths $U$,
$V_{1,2}$ and $J_1^z$ in such a way that \fr{finetunings} are achieved.
In order to ascertain the low-energy properties of \fr{Hext}, we
compute the energies of the ground state and several low-lying excited
states for a quarter-filled band, and compare the results to
expectations based on Luttinger liquid theory. We choose to work at
quarter filling in order to simplify finite-size scaling analyses. As
alluded to in Appendix \ref{app:irrops}, working at commensurate
fillings induces additional Umklapp interactions. In the case at hand
this corresponds to the presence of an additional perturbation $\int
dx\ \cos (2 \Phi_c)$. However this term has scaling dimension $8K_c$
and is therefore highly irrelevant for $K_c\geq 1/2$. We therefore
discard it in the following analysis.


Some insight into how the parameters $K_c,K_s$ and $g_{s,1}$ depend on
$U$, $V_{1,2}$ and $J_1^z$ can be gained by bosonizing the
interactions at weak coupling. Using (\ref{opexpansion}) we obtain
in leading order 

\begin{eqnarray}
n_{j,\uparrow}n_{j,\downarrow} &\sim& \frac{{\cal A}}{4} 
\Big\lbrack
\mc O_4^{(2)}(x)+2 \mc O_5^{(2)}(x)- \mc O_2^{(2)}(x)\nn
&&-2 \mc O_3^{(2)}(x) +8\eta_\uparrow \bar{\eta}_\downarrow\eta_\downarrow
\bar{\eta}_\uparrow
{\cal O}^{(2)}_1(x) \Big\rbrack,  \label{opeU}\\
n_jn_{j+1}& \sim  & {\cal A} \left\lbrack \mc O_4^{(2)}(x)+2
\mc O_5^{(2)}(x)\right\rbrack, \label{opeV1} \\
n_jn_{j+2}& \sim &\frac{{\cal A}}{2} \Big \lbrack3 \mc O_4^{(2)}(x)+ 6\mc O_5^{(2)}(x) +
\mc O_2^{(2)}(x)\nonumber \\ 
& & + 2 \mc O_3^{(2)}(x) -8\eta_\uparrow \bar{\eta}_\downarrow\eta_\downarrow
\bar{\eta}_\uparrow
{\cal O}^{(2)}_1(x) \Big\rbrack,\label{opeV2} \\
S_j^zS_{j+1}^z&\sim &\frac{{\cal A}}{4}\left\lbrack \mc O_2^{(2)}(x)+2
\mc O_3^{(2)}(x)\right\rbrack.
\label{opeJz}
\end{eqnarray}
Here $x=j a_0$, ${\cal A}$ is a dimensionful amplitude, and
\bea
{\cal O}^{(2)}_1&=&\cos\Phi_s\ ,\nn
\mc O_{2}^{(2)}&=&(\partial_x\varphi_s)^2 +
(\partial_x\bar{\varphi}_s)^2\ ,\nn
\mc O_{3}^{(2)}&=&\partial_x\varphi_s \partial_x\bar{\varphi}_s\ ,\nn
\mc O_{4}^{(2)}&=&(\partial_x\varphi_c)^2 +
(\partial_x\bar{\varphi}_c)^2\ ,\nn
\mc O_{5}^{(2)}&=& \partial_x\varphi_s \partial_x\bar{\varphi}_s.
\eea
At weak coupling the spin-charge separated Luttinger liquid at low
energies is therefore perturbed by 
\be
\delta H=\sum_{j=1}^5\nu_j\int dx\ {\cal O}^{(2)}_j(x)\equiv
\sum_{j=1}^5\delta H_j,
\ee
where the $\nu_j$ are proportional to linear combinations of
$U$, $V_1$, $V_2$ and $J^z_1$. The effects of $\delta H_2$ and $\delta
H_4$ are to renormalize the spin and charge velocities respectively, while
$\delta H_3$ and $\delta H_5$ change the values of the Luttinger parameters
$K_s$ and $K_c$. We must pay special attention to the perturbation
$\delta H_1$: as discussed in Appendix \ref{app:irrLE}, this operator
is marginal at weak coupling, but gives rise to the relevant spinon
mass term  as we approach the LE point. Our objective is to adjust
  $U$, $V_1$, $V_2$ 
and $J^z_1$ in such a way that $K_{c,s}$ are reduced towards $1/2$,
while the coupling $\nu_1$ of the spinon mass term remains very
small. The bosonization results \fr{opeJz} suggest the following
prescription for achieving this at weak coupling:
\begin{itemize}
\item{} Increase $V_2$ in order to make $|\nu_1|$ very small;
\item{} Then adjust $V_1$ and $J^z_1$ in order to drive $K_{c,s}$
towards $1/2$. Importantly, the bosonization results \fr{opeJz}
indicate that at least at weak coupling this does not produce sizeable
contributions to $\delta H_1$.
\end{itemize}
Our analysis below is guided by these considerations, even though we
are not operating in the weak coupling regime, in which \fr{opeJz} are
applicable. 

\subsection{Finite-size spectrum of unperturbed Luttinger liquid with
  open boundary conditions}
\label{sec:Ksandvs}
A standard procedure for determining the Luttinger parameters $K_c$
and $K_s$ is to compare  the finite-size spectrum predicted by
Luttinger liquid theory with the low-energy spectrum calculated
numerically for a given lattice model.  The finite-size spectrum 
relative to the ground state of a Luttinger liquid with open
boundaries is given by
\bea
&&\Delta E(S^z,\Delta N_c, \{m_{\ell}^c, m^s_{\ell}\})=\frac{\pi v_s
  (S^z)^2}{K_s L } +\frac{\pi v_c  (\Delta N_c)^2}{4 K_c L}\nn
&&\qquad-\frac{\pi v_c d \Delta N_c}{2K_c L}   +\sum_{\alpha=c,s}\sum_{\ell=1}^\infty\frac{\pi v_\alpha}{L} \ell m^\alpha_{\ell}+o(L^{-1}).\label{espectrum}
\eea
Here $S^z$ and $\Delta N_c$ are quantum numbers  in the spin and
charge sectors, respectively, associated with   the global
U(1)$\otimes$U(1) symmetry.  The value of $S^z$ measures the change in
the total magnetization and  $\Delta N_c=N-N_0$ measures the change in
the total number of electrons with respect to a singlet ground state
with $N_0$ electrons. As usual, the  values of $S^z$ and $\Delta N_c$
are constrained by the selection rule that $2S^z$ must be even (odd)
if $\Delta N_c$ is even (odd). The parameter $d$ in
Eq. (\ref{espectrum}) is a dimensionless constant that depends on the
definition of the chemical potential   to order $1/L$.  The parameters
$m_\ell^s$ and $m_\ell^c$ are non-negative integers that count the
number of low-energy particle-hole pairs in each sector. We note that
for open boundary conditions the total momentum is  not conserved and
there are no quantum numbers associated with current excitations
(i.e. with transferring particles between Fermi points). Thus
(\ref{espectrum}) should not be confused with the spectrum in
(\ref{EMIM2}), which is valid for periodic boundary conditions.  

The spin and charge velocities $v_s$ and $v_c$ can be extracted by
matching the energies of the lowest excitations with $S^z=\Delta
N_c=0$. Assuming $v_s<v_c$ (which can be verified  by analyzing
excitations with different quantum numbers), the first  excited state
corresponds to $m_1^s=1$ and has  energy
\be 
\Delta E(0,0,\{m_\ell^s=\delta_{\ell,1},m^c_\ell=0\})=\frac{\pi v_s}{L}. \label{firstspinexc}
\ee
Likewise, if $v_s<v_c<2v_s$, the second excited state corresponds to $m_1^c=1$ and has energy \be
\Delta E(0,0,\{m_\ell^s=0,m^c_\ell=\delta_{\ell,1}\})=\frac{\pi v_c}{L}. \label{firstchargeexc}
\ee

Having determined the velocities, one can obtain the Luttinger
parameters by analyzing low-lying excitations that change the quantum
numbers $S^z$ and $\Delta N_c$. We adopt the short-hand notations 
\be
\Delta E_0(S^z,\Delta N_c)\equiv \Delta E(S^z,\Delta N_c,\{m_\ell^c=m_\ell^s=0\,\,\forall \ell\})
\ee
for the lowest energies in each sector of fixed $S^z$ and $\Delta
N_c$.  To isolate the dependence on $K_c$, we consider excitations
with $\Delta N_c=\pm 2$ and $S^z=0$. The dependence on the unknown
constant $d$ can be eliminated by taking the combination
\be
\frac{L}{4}[\Delta E_0(0,2)+\Delta E_0(0,-2)]\simeq\frac{\pi v_c}{2K_c}=\kappa^{-1},\label{compressib}
\ee
where we recognize $\kappa$ as the compressibility of the   Luttinger
liquid\cite{giamarchi}. Analogously, $K_s$ can be determined using the
relation for the finite-size spin gap
\be 
\frac{L}{2}\Delta E_0(1,0)=\frac{\pi v_s}{2K_s}.
\label{suscept}
\ee
The right-hand side of Eq. (\ref{suscept}) is equal to the inverse
spin susceptibility.  

The procedure described above is standard for Luttinger liquids
that are only perturbed by (strongly) irrelevant operators. In our
case, however, we must consider the effects of the \emph{relevant}
operator $\cos\Phi_s$, which generates the spinon mass, as well as the
interaction $\partial_x\Phi_c\cos\Phi_s$, which is
marginal at the LE point and only weakly irrelevant in its
vicinity. These perturbations introduce corrections to
(\ref{firstspinexc}), (\ref{firstchargeexc}), (\ref{compressib}) and
(\ref{suscept}), which affect the finite-size scaling analysis. Since 
$\cos\Phi_s$ is the leading perturbation, we first focus our efforts on
fine tuning it to zero, as we discuss in the next subsection.


\subsection{Fine tuning the spinon mass}


As anticipated in Sec. \ref{sec:latticemodel}, our
strategy for fine tuning the spinon mass to zero starts by
suppressing the coupling constant of the perturbation $\delta H_1$.
According to (\ref{opeV2}), for fixed $U>0$  this can be done
efficiently  by increasing the next-nearest-neighbor interaction
$V_2>0$. In this 
process, we keep $V_1=J_1^z=0$, so the SU(2) symmetry is preserved. As
we keep increasing $V_2$, the marginal coupling constant will change
sign at some critical value $V_2^c$. Beyond  this point the
perturbation becomes marginally relevant and the system undergoes a
Berezinskii-Kosterlitz-Thouless (BKT) transition to a spin-gapped
phase, analogous to the dimerization transition in the $J_1$-$J_2$
spin chain~\cite{Eggert1996}.  

Because of the exponentially small value of the gap in the vicinity of
the critical point it is difficult to pinpoint $V_2^c$ by means of a gap
scaling analysis. A better approach is to determine the critical point by 
searching for a level crossing in the spin excitation spectrum in the
neutral sector $S^z=\Delta N_c=0$ \cite{Nomura}. The idea is that in a
Luttinger liquid the state  with $m_2^s=1$ should be degenerate with
the state $m_1^s=2$; however, in practice the degeneracy is lifted at
order $(L\ln L)^{-1}$ due to the marginal perturbation. Exactly at the
critical point, the coupling constant for $\cos(\Phi_s)$ vanishes, and
$V_2^c$ can be identified from the level crossing between $m_2^s=1$
and $m_1^s=2$ states.  

\subsection{Analysis of finite-size excitation energies}
\label{ssec:FSE}

After tuning the   coupling constant of $\delta H_1$ to zero, we
proceed to adjusting  $V_1$,  and $J_1^z$ in order to achieve 
values $K_{c,s}\approx \frac{1}{2}$. This has to be done 
while making sure that the coupling constant $g_{s,1}$ in \fr{Hcs}, or
equivalently $\lambda_1$ in \fr{luttinger}, is held at zero as
we approach the LE point.  To determine the values of 
$K_{c,s}$, and also to ascertain that we remain in the Luttinger
liquid phase as we vary the parameters of the lattice model, we resort
to an analysis of the finite size energy levels of the ground state
and low-lying excitations. For this purpose, we now refine the
expressions presented in  Sec. \ref{sec:Ksandvs} by including the
effects of the leading perturbations in the vicinity of the LE point.

An inherent difficulty encountered when approaching the LE point is
that corrections to the Luttinger liquid form \fr{espectrum} of the
finite-size energy spectrum become increasingly complicated. To see
this let us consider the bosonized form \fr{luttinger} of our
Hamiltonian. 
In the regime $\frac{1}{2}<K_{s,c}<1$ we expect the structure of the
energy difference $\Delta E_0(S^z,0)$  to be of the form
\begin{eqnarray}
\Delta E_0(S^z,0)&=&  \frac{a_1}{\tilde{L}^{2K_s-1}}+\frac{\pi v_s
  (S^z)^2}{K_s \tilde{L}}+\frac{b_1}{\tilde{L}^{2K_s}}\nn
&&+\frac{b_2}{\tilde{L}^{4K_s-1}}+\frac{b_3}{\tilde{L}^{6K_s-2}}
+ \ldots,
\label{fitE0Sz}
 \end{eqnarray}
where we have defined $\tilde{L}=L+1$. The origin of the various
contributions is as follows:
\begin{itemize}
\item The $a_1$ term arises from first order perturbation theory in
the spinon mass term $\lambda_1\int dx\ \cos\Phi_s$. As a consequence
of our fine-tuning we expect $a_1$ to be quite small, so that higher
orders of perturbation theory can be neglected.
\item The $b_1$ term arises from first-order perturbation theory in
$\lambda_2\int dx\ \partial_x\Phi_c\cos\Phi_s$ (which generally is
non-zero for open boundary conditions).
\item{} The $b_2$ and $b_3$ terms arise from second- and
third-order perturbation theory in $\lambda_2\int
dx\ \partial_x\Phi_c\cos\Phi_s$ respectively. Here we need to consider
higher orders in perturbation theory because the bare $\lambda_2$ has
not been fine tuned and is not guaranteed to be small.
\end{itemize}
We see that the size dependence becomes increasingly complex as $K_s$
tends towards $1/2$, which complicates the analysis of energy
levels. A good way of achieving an accurate description of finite-size
energies would be through (two-loop) renormalization-group-improved
perturbation theory around both the LE point and the weak-coupling
limits. However, as this is quite involved, we content ourselves with
the simpler form \fr{fitE0Sz} (and stress again that our numerical
analysis is to be considered as preliminary).

\subsubsection{Determining the Luttinger parameter \texorpdfstring{$K_s$}{LG}}
\label{ssec:Ks}
Equation~(\ref{fitE0Sz}) provides us with an alternative way of
determining $K_s$ by considering the system-size dependence of the
quantity
\begin{eqnarray}
\label{defS}
\delta(L) \equiv \frac{1}{4} \Delta E_0(2,0)-\Delta E_0(1,0).
\end{eqnarray} 
Using the expression \fr{fitE0Sz} we obtain
\begin{eqnarray}
\delta(L)&=&
\frac{\tilde{a}_1}{\tilde{L}^{2K_s-1}}+\frac{\tilde{b}_1}{\tilde{L}^{2K_s}}+\frac{\tilde{b}_2}{\tilde{L}^{4K_s-1}}+\ldots.
\label{fitS}
\end{eqnarray}
By fitting $\delta(L)$ computed numerically for a range of system
sizes to \fr{fitS}, we obtain a value for $K_s$.

\subsubsection{Determining the spin velocity 
\texorpdfstring{$v_s$}{LG}}
Given $K_s$, we may determine $v_s$ from the size dependence   in
\fr{fitE0Sz}. Alternatively we can consider other excited states,
whose finite-size energies can be analyzed similarly. For example, for
$v_s<v_c$, the size dependence of the first excited state in the
neutral sector is given by 
\begin{eqnarray}
&&\Delta E(0,0,\{m_\ell^s=\delta_{\ell,1},m^c_\ell=0\}) \nn
&&\quad=\frac{\tilde{a}_1^s}{\tilde{L}^{2K_s-1}}+\frac{\pi v_s
}{\tilde{L}}
+\frac{\tilde{b}_1^s}{\tilde{L}^{2K_s}}
+\frac{\tilde{b}_2^s}{\tilde{L}^{4K_s-1}}+\ldots
\label{fitvs} 
\end{eqnarray}
By computing the energy difference \fr{fitvs} numerically for a range
of system sizes, using $\tilde{a}_1^s$ and $\tilde{b}_{1,2}^s$ as fit
parameters, and fixing $K_s$ to be the value obtained from the
analysis of $\delta(L)$ in Eq. (\ref{fitS}), we may extract the spin velocity $v_s$.

\subsubsection{Determining the charge velocity 
\texorpdfstring{$v_c$}{LG}}
\label{ssec:vc}
The structure of finite-size corrections to energy levels of charge
excitations is somewhat simpler because some of the terms in first-order perturbation theory vanish. For $v_s<v_c<2v_s$, the energy of
the second excited state in the neutral sector scales as
\begin{eqnarray}
&&\Delta E(0,0,\{m_\ell^s=0,m^c_\ell=\delta_{\ell,1}\}) \nn
&&\qquad\qquad
=\frac{\pi v_c
}{\tilde{L}}+\frac{\tilde{b}_2^c}{\tilde{L}^{4K_s-1}}+\ldots
\label{fitE2}
\end{eqnarray}
Using the results for $K_s$ from the analysis described in
\ref{ssec:Ks}, Eq.~\fr{fitE2} provides us with a means of determining
$v_c$ through a finite-size scaling analysis.
\subsubsection{Determining \texorpdfstring{$K_c$}{LG}}
Finally, the energy difference related to the
compressibility \fr{compressib} is found to have the following size
dependence 
\begin{eqnarray}
\frac{\Delta E_0(0,2)+\Delta E_0(0,-2)}{16}&=&\frac{\pi v_c}{8 K_c
  \tilde{L}}+\frac{b_2^c}{\tilde{L}^{4K_s-1}}+\ldots\nn 
&\equiv&f({L}).
\label{fitkappa}
\end{eqnarray}
We can use \fr{fitkappa} together with $K_s$ and $v_c$ determined in
sections \ref{ssec:Ks} and \ref{ssec:vc} respectively, to fix the
Luttinger parameter $K_c$ through a finite-size scaling analysis of
$f({L})$, taking ${b}_2^c$ to be a fit parameter.

\subsection{Numerical results}
We now turn to the numerical implementation of the method set out in
the previous subsection. We performed DMRG\cite{DMRG,DMRG2}
computations on lattices with up to $L = 216$ 
sites, keeping up to 3000 states and running up to 36 finite-size
sweeps. In case of the Hubbard model we have checked the energies
obtained in this way against exact Bethe Ansatz results and found the
relative errors to be of order $10^{-9}$.

\begin{figure}[ht]
\includegraphics[scale=0.3]{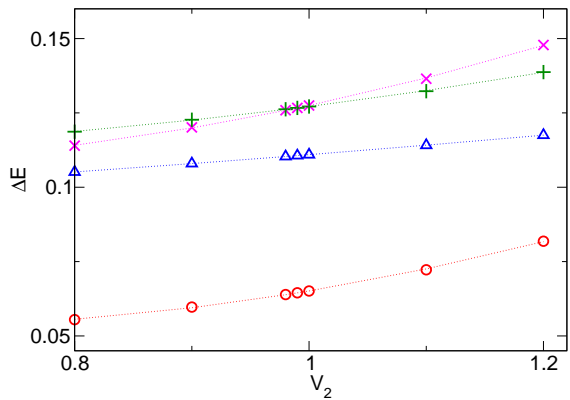}
\caption{Energies of the four lowest excited states in the neutral
sector $S^z=\Delta N_c=0$ for $U=3$, $J^z_1=V_1=0$ and $L = 64$ as a
function of $V_2$. The critical value $V_2^c$ is identified through
the crossing of the two nearly degenerate states in the $(2s,0)$
multiplet. Lines are guide to the eye.}
\label{levelcrossing} 
\end{figure}

We work with model (\ref{Hext}) at fixed $U=3$ and search for the LE
point by varying $V_1$, $V_2$ and $J_1^z$. The first step is to
increase $V_2$ keeping $V_1=J_1^z=0$. In Fig.~\ref{levelcrossing} we
show the first four excitations in the neutral sector for  $L=64$ as a
function of $V_2$. By tracking the evolution of the energies starting
from $V_2=0$, i.e. the Hubbard model, we are able to identify the
quantum numbers $\{m^s_\ell , m^c_\ell\}$ for each of these 
states. The two nearly degenerate spin descendant states of interest
correspond to the third and fourth excited states. From the crossing
of these two energy levels  we estimate $V_2^c\approx 1.1$. 
 
Next, we vary $V_1$ and $J_1^z$ to bring $K_{c}$ and $K_s$ close to
$1/2$. After searching in parameter space, we settle  for the
particular  parameter set
\be
U=3\ ,\ V_1=0.85\ ,\ V_2=1.1\ ,\ J_1^z=5.2.
\label{parametervalues}
\ee
The analysis presented below
suggests that this corresponds to Luttinger parameter values of
$K_s=0.655$ and $K_c=0.500$. This is probably as close to the LE point
as one can get without a more precise theoretical description of
finite-size energies based on a renormalization-group-improved
perturbation theory analysis.

\begin{figure}[ht]
\includegraphics[scale=0.3]{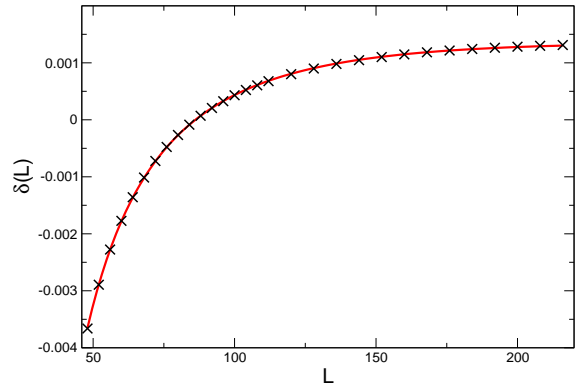}
\caption{Crosses: numerical results for $\delta(L)$ as defined in
Eq.~(\ref{defS}). 
The solid line is the best fit to Eq.~\fr{fitS} and is obtained for
$K_s=0.655, \tilde{a}_1=0.009$, $\tilde{b}_1=0.493$ and
$\tilde{b}_2=-5.131$.} 
\label{fig_S}
\end{figure}

Figure~\ref{fig_S} shows numerical results for the quantity $\delta(L)$
defined in Eq.~(\ref{defS}).
The data have been fitted using (\ref{fitS}). The resulting best-fit
estimates are
\be
K_s=0.655,
\ee
$\tilde{a}_1=0.009$, $\tilde{b}_1=0.493$ and
$\tilde{b}_2=-5.131$. The quality of the fit is visibly excellent (the
residuals are $\sim10^{-6}$). The numerical value for $\tilde{a}_1$ is
very small, confirming that we have almost succeeded with fine-tuning
the spinon mass term to zero (on the scale set by the system sizes we
consider).

In the next step we determine the spin velocity using (\ref{fitE0Sz})
and retaining the $b_1$ and $b_2$ terms. Setting $K_s=0.655$ we obtain
\be
v_s\approx 1.707.
\ee 
This value is consistent with the result obtained by considering
$\Delta E(0,0,\{m_\ell^s=\delta_{\ell,1},m^c_\ell=0\})$ in Eq. (\ref{fitvs}). 
We have verified that the energy level corresponds to the first spin
descendant state by tracking the tower of lowest-lying energies in the
neutral sector along a path in parameter space connecting the Hubbard
model to the point \fr{parametervalues}.

\begin{figure}
\includegraphics[scale=0.3]{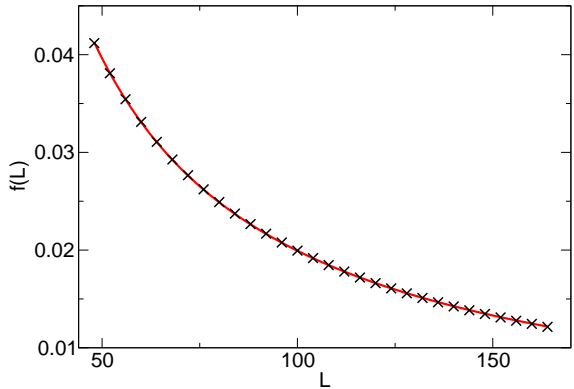}
\caption{Crosses: numerical results for $f(L)$ as defined in
  (\ref{fitkappa}). The solid line is a best fit to \fr{fitkappa} 
with $K_s=0.655$.}
\label{fig_kappa}
\end{figure}

Having determined $v_s$ and $K_s$, we now turn to the charge sector.
In Fig.~\ref{fig_kappa} we present numerical results for the quantity
$f(L)$ defined in Eq. \fr{fitkappa}, together with a fit to the
functional form posited in \fr{fitkappa}. The spin Luttinger parameter
is fixed as $K_s=0.655$. The best-fit estimates are 
\be
\frac{\pi v_c}{8 K_c}=2.000,
\label{vckc}
\ee
and $b_2^c=0.271$. The residuals are of order $\mathcal{O}(10^{-5})$. 

\begin{figure}
\includegraphics[scale=0.3]{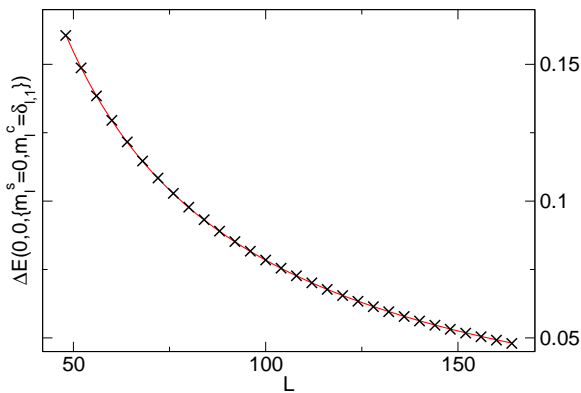}
\caption{Crosses: numerical results for $\Delta
E(0,0,\{m_\ell^s=0,m^c_\ell=\delta_{\ell,1}\})$. The solid line
is a best fit to \fr{fitE2} for $K_s=0.655$.
} 
\label{fig_vc}
\end{figure}

Finally, in Fig.~\ref{fig_vc} we present numerical results
for $\Delta E(0,0,\{m_\ell^s=0,m^c_\ell=\delta_{\ell,1}\})$. Fitting
the data to \fr{fitE2} with $K_s=0.655$ results in estimates
\be
v_c\approx 2.543,
\label{fitvc}
\ee
and $\tilde{b}_2^c=-1.191$. 
Combining \fr{vckc} and \fr{fitvc} we conclude that
\be
K_c\approx0.500.
\ee

\subsection{Friedel oscillations}
The analysis of finite-size energy levels presented above is clearly
rather involved, and an independent check on the results for $K_{c,s}$
would clearly be very useful. Such a check is provided by analyzing
Friedel oscillations of the charge density on a chain with open
boundary conditions, cf Ref.~\onlinecite{soeffing2009}. We summarize
the main steps of how to calculate the charge density for open
boundary conditions in (perturbed) Luttinger liquid theory in
Appendix~\ref{app:parity}. For a quarter filled band we obtain
\bea
n_j&\approx&\frac{1}{2}+\frac{d_1}{(D_j)^{2 K_s}}+
d_2
\frac{\sin\big(\big\lbrack\frac{\pi}{2}
+\frac{p_1}{L+1}\big\rbrack j +p_2\big)}
{(D_j)^{(K_c+K_s)/2}}\nonumber \\&& +d_3
\frac{\cos\big(\big\lbrack\pi+\frac{2p_1}{L+1}\big\rbrack j +2 p_2
  \big)}{(D_j)^{2 K_c}}+\ldots,
\label{eq:dens0}
\eea
where $D_j$ denotes
\be
D_j=\frac{2(L+1)}{\pi}\sin\big(\frac{\pi j}{L+1}\big).
\ee

\begin{figure}
\includegraphics[scale=0.3]{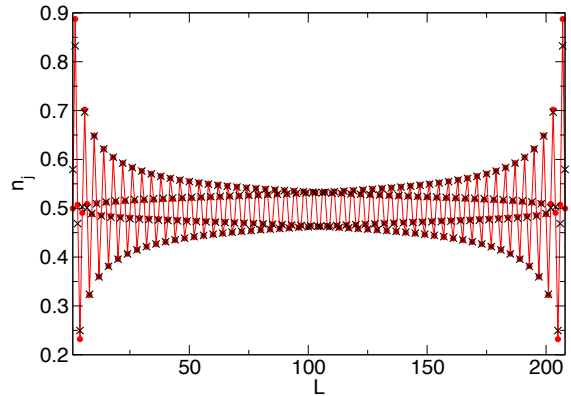}
\caption{Crosses: local density for a quarter filled band and
$U=3$, $V_1=0.85$, $V_2=1.1$, $J_1^z=5.2$. The solid
  line is a fit to eqn (\ref{eq:dens0}) with $K_s=0.65$ and $K_c=0.5$.}
\label{fig:dens0}
\end{figure}

In Fig.~\ref{fig:dens0} we compare the prediction \fr{eq:dens0} to
DMRG results for a quarter filled band and system size $L=208$. We fix
the values of the Luttinger parameters to $K_s=0.65$ and $K_c=0.5$ and
use the amplitudes $d_j$ and phase shifts $p_j$ as fit parameters. The
agreement is very good except near the boundaries.  
The best fit is obtained for $d_1=-0.05$, $d_2=0.83$, $d_3=0.01$, 
$p_1=-\pi/4$ and $p_2=-\pi/2$. 
The good agreement between the expected behaviour \fr{eq:dens0} and
the DMRG results provides a consistency check on the values of
$K_{c,s}$ extracted from the analysis of finite-size energy levels.

\section{Summary and Conclusions}
In this work we have developed a new approach to deriving mobile 
impurity models for studying dynamical correlations in gapless models
of spinful fermions. Our construction is based on the principle  that
our mobile impurity must carry the quantum numbers of a
holon/antiholon (charge $\pm e$, spin $0$) or a spinon (charge $0$ and
spin $\pm 1/2$). For the case of integrable models like the Hubbard
chain, it is known from the exact solution  that elementary excitations
at finite energies are (anti)holons and spinons with precisely these
quantum numbers. In integrable models these excitations are stable
(i.e. do not decay). Breaking integrability is expected to render
their lifetimes finite, but leaving their spin and charge quantum
numbers intact. 

To facilitate our construction, we first derived a representation of
spinful nonlinear Luttinger liquids in terms of strongly interacting
fermionic holons and spinons. At a particular \emph{Luther-Emery}
point for spin \emph{and} charge, holons and spinons become
noninteracting. Using this as our point of reference, we derived a
mobile impurity model appropriate for the description of threshold
singularities in the single-particle spectral function for a general
class of extended Hubbard models in their Luttinger liquid phase.

Our construction differs in important aspects from previous work by
Schmidt, Imambekov and Glazman\cite{SIG1,SIG2}. However, we 
demonstrated explicitly how and why results for threshold exponents
obtained in the two approaches coincide.

Finally, we presented a preliminary analysis of the question of how to
realize, in the low-energy regime, the Luther-Emery point for spin and
charge in a lattice model of interacting spinful fermions.
We showed that the structure of allowed perturbations to the
Luther-Emery point is such that fine tunings of various interactions
is required. Achieving these fine tunings is very delicate, and we
discussed in some detail what problems one encounters.

Our work raises a number of interesting questions that deserve further
attention. First and foremost, further numerical studies are required
in order to identify a parameter regime in an appropriate extended
Hubbard model that, at least approximately, realizes the Luther-Emery
point. Second, it would be very interesting to implement the numerical
procedure we proposed for determining the parameters $\gamma_\alpha$,
$\bar\gamma_\alpha$ characterizing the threshold exponents. One first
might want to reproduce the known exact results for the Hubbard model,
before moving on to non-integrable cases. Third, in close proximity to
the Luther-Emery point the mobile impurity model involves an additional
marginal interaction that cannot be removed by a unitary
transformation. It would be interesting to analyze its effects on
the threshold exponents.

\acknowledgments
This work was supported by the EPSRC under grants EP/I032487/1 (FHLE)
and EP/J014885/1 (FHLE), by the IRSES network QICFT (RP and FHLE), 
the CNPq (RGP), and by the DFG under grant SCHN 1169/2-1 (IS). We are
grateful to Eric Jeckelmann for providing the DMRG code used in the
numerical part of our analysis.
\appendix

\section{Irrelevant perturbations to the Luttinger liquid Hamiltonian}
\label{app:irrops}
One way of working out the allowed irrelevant perturbations to the
Luttinger liquid Hamiltonian is by using symmetry considerations. 
Our starting point are extended Hubbard models of the kind
\fr{HEHubb}. For the purposes of this appendix, we will assume all
interactions to be \emph{small}. 

\subsection{Symmetries} 
Our lattice models of interest are invariant under various symmetry
operations. These symmetries are inherited by the bosonic low-energy
description \fr{luttinger} and we now discuss their realizations.

\subsubsection{Spin flip symmetry} The lattice models of interest are
invariant under exchange of up and down spins
\be
c_{j,\up}\leftrightarrow c_{j,\down}.
\ee
is realized at the level of the bosonic fields as
\be
\begin{pmatrix}
\varphi_s(x)\\
\bar\varphi_s(x)\\
\eta_\up\\
\bar\eta_\up
\end{pmatrix}
\longrightarrow -
\begin{pmatrix}
\varphi_s(x)\\
\bar\varphi_s(x)\\
\eta_\down\\
\bar\eta_\down
\end{pmatrix}.
\ee
\subsubsection{Translational Invariance:}
The translation operator acts as
\be
T^\dagger c_{j,\sigma}T=c_{j+1,\sigma}.
\ee
We can implement this on the level of bosonic fields by imposing the
transformation properties
\bea
\begin{pmatrix}
\varphi_c(x)\\
\bar\varphi_c(x)
\end{pmatrix}
&\longrightarrow&
\begin{pmatrix}
\varphi_c(x)-2k_Fa_0\\
\bar\varphi_c(x)-2k_Fa_0
\end{pmatrix}\ ,\nn
\begin{pmatrix}
\varphi_s(x)\\
\bar\varphi_s(x)
\end{pmatrix}
&\longrightarrow&
\begin{pmatrix}
\varphi_s(x)\\
\bar\varphi_s(x)
\end{pmatrix}.
\eea
We note that this transformation works for all higher harmonics in
\fr{opexpansion}. Translational invariance of the Hamiltonian then implies that
it generically may not contain any vertex operators of the charge
boson. Exceptions to this rule occur at \emph{commensurate fillings}
\be
k_Fa_0=\pi \frac{p}{q}\ ,\quad p,q\in\mathbb{N}.
\ee
Here operators of the form
\be
\cos\big(\frac{q}{2}\Phi_c\big)\ ,\quad \sin\big(\frac{q}{2}\Phi_c\big)
\ee
are allowed to occur. As long as $q> 4$ such operators are highly
irrelevant and do not play a role in the following discussion.
\subsubsection{\texorpdfstring{$U(1)\otimes U(1)$}{LG} Invariance}
By this mean we that the Hamiltonian commutes with particle number and
the z-component of total spin
\be
[H,\hat{S}^z]=[H,\hat{N}]=0.
\ee
This symmetry implies that no vertex operators involving the dual
fields $\Theta_c$, $\Theta_s$ are allowed to occur in the expression
for ${\cal H}$.
\subsubsection{Site Parity}
The reflection symmetry acts on the lattice fermion operators like
\be
Pc_{j,\sigma}P=c_{-j,\sigma}.
\ee
We see that $P$ can be realized in the field theory as
\bea
P\varphi_a(x)P&=&-\bar{\varphi}_a(-x)\ ,\quad
P\bar\varphi_a(x)P=-{\varphi}_a(-x)\ ,\nn
P\eta_\sigma P&=&\bar{\eta}_\sigma.
\label{siteparity}
\eea
Crucially, parity acts on the unphysical Klein degrees of freedom as
well. Moreover, we obtain the following constraints on the amplitudes
$A_{n,m}$ in \fr{opexpansion}
\be
A_{n,m}=A_{1-n,-1-m}.
\ee
\subsection{Dimension two operators allowed by symmetry}
We now list all symmetry-allowed perturbations to the Luttinger liquid
Hamiltonian for noninteracting spinful fermions with scaling
dimensions $2$, $3$ and $4$. Such perturbations will be of the form
\be
\sum_j \hat\Gamma_j\lambda_j\int dx\ {\cal O}(x),
\ee
where $\lambda_j$ are coupling constants and $\Gamma_j$ are products
of Klein factors. The only non-trivial combination of Klein factors
allowed to appear is in fact
\be
\eta_\up\bar\eta_\up\eta_\down\bar\eta_\down.
\ee
This is because the all terms in our Hamiltonian \fr{HEHubb} are of
the form
\be
c^\dagger_{j,\sigma}c_{k,\sigma},\quad
c^\dagger_{j,\sigma}c^\dagger_{k,\tau}c_{l,\tau}c_{m,\sigma}\ .
\ee
Expressing the lattice fermion operators in terms of Bose fields by
\fr{opexpansion}, and then imposing that for a given interaction to
appear in the Hamiltonian it must not contain a rapidly oscillating
factor $e^{ijk_Fx }$, one finds that the Klein factors either cancel
or combine to $\eta_\up\bar\eta_\up\eta_\down\bar\eta_\down$.

Taking this into account, we find five symmetry-allowed dimension two
operators  
\bea
{\cal O}^{(2)}_1&=&\cos\Phi_s\ ,\nn
{\cal O}^{(2)}_2&=&(\partial_x\varphi_s)^2+(\partial_x\bar\varphi_s)^2\ ,\nn
{\cal O}^{(2)}_3&=&\partial_x\varphi_s\partial_x\bar\varphi_s\ ,\nn
{\cal O}^{(2)}_4&=&(\partial_x\varphi_c)^2+(\partial_x\bar\varphi_c)^2\ ,\nn
{\cal O}^{(2)}_5&=&\partial_x\varphi_c\partial_x\bar\varphi_c\ .
\eea
We note that none of these leads to a coupling between spin and charge
sectors. In order to see that the operator $\cos\Phi_s$ is allowed, but
$\sin\Phi_s$ is not, one needs to consider the structure of Klein
factors in \fr{opexpansion}.
\subsection{Dimension three operators allowed by symmetry}
The analogous analysis for dimension three operators gives the
three possible perturbations involving only the charge sector
\bea
{\cal O}^{(3)}_1&=&(\partial_x\varphi_c)^3+(\partial_x\bar\varphi_c)^3\ ,\nn
{\cal O}^{(3)}_2&=&(\partial_x\varphi_c)^2\partial_x\bar\varphi_c
+(\partial_x\bar\varphi_c)^2\partial_x\varphi_c\ ,\nn
{\cal O}^{(3)}_3&=&\partial_x\varphi_c\partial^2_x\bar
\varphi_c-\partial_x\bar \varphi_c\partial^2_x \varphi_c,
\eea
and four possible perturbations that couple spin and charge sectors
together 
\bea
{\cal O}^{(3)}_4&=&(\partial_x\varphi_s)^2\partial_x\varphi_c
+(\partial_x\bar\varphi_s)^2\partial_x\bar\varphi_c\ ,\nn
{\cal O}^{(3)}_5&=&(\partial_x\varphi_s)^2\partial_x\bar\varphi_c
+(\partial_x\bar\varphi_s)^2\partial_x\varphi_c\ ,\nn
{\cal O}^{(3)}_6&=&\partial_x\varphi_s\partial_x\bar\varphi_s
(\partial_x\bar\varphi_c+\partial_x\varphi_c)\ ,\nn
{\cal
  O}^{(3)}_7&=&\big(\partial_x\varphi_c+\partial_x\bar\varphi_c\big)
\cos(\Phi_s)\ .
\eea

\subsection{Dimension four operators allowed by symmetry}
Finally we consider all symmetry-allowed dimension four operators. 

\subsubsection{Charge sector only}
We find six perturbations involving only the charge sector
\bea
{\cal O}^{(4)}_1&=&(\partial_x\varphi_c)^4+(\partial_x\bar\varphi_c)^4\ ,\nn
{\cal O}^{(4)}_2&=&(\partial_x\varphi_c)^2(\partial_x\bar\varphi_c)^2\ ,\nn
{\cal O}^{(4)}_3&=&(\partial_x\varphi_c)^3\partial_x\bar\varphi_c
+(\partial_x\bar\varphi_c)^3\partial_x\varphi_c\ ,\nn
{\cal O}^{(4)}_4&=&(\partial_x^2\varphi_c)^2+(\partial_x^2\bar\varphi_c)^2\ ,\nn
{\cal O}^{(4)}_5&=&\partial_x^2\varphi_c\partial_x^2\bar\varphi_c\ ,\nn
{\cal O}^{(4)}_6&=&\partial_x^2\varphi_c(\partial_x\bar\varphi_c)^2-
\partial_x^2\bar\varphi_c(\partial_x\varphi_c)^2.
\eea
Here we have used that 
\bea
\partial_x^3\varphi_c\partial_x\bar\varphi_c+
\partial_x^3\bar\varphi_c\partial_x\varphi_c&=&
-2\partial_x^2\varphi_c\partial_x^2\bar\varphi_c\nn
&+&\text{total derivative}.
\eea
\subsubsection{Spin sector only}
We find altogether eight symmetry allowed perturbations involving only
the spin sector
\bea
{\cal O}^{(4)}_7&=&(\partial_x\varphi_s)^4+(\partial_x\bar\varphi_s)^4\ ,\nn
{\cal O}^{(4)}_8&=&(\partial_x\varphi_s)^2(\partial_x\bar\varphi_s)^2\ ,\nn
{\cal O}^{(4)}_9&=&(\partial_x\varphi_s)^3\partial_x\bar\varphi_s
+(\partial_x\bar\varphi_s)^3\partial_x\varphi_s\ ,\nn
{\cal O}^{(4)}_{10}&=&(\partial_x^2\varphi_s)^2+(\partial_x^2\bar\varphi_s)^2\ ,\nn
{\cal O}^{(4)}_{11}&=&\partial_x^2\varphi_s\partial_x^2\bar\varphi_s\ ,\nn
{\cal O}^{(4)}_{12}&=&\cos(\Phi_s)\partial_x\varphi_s\partial_x\bar\varphi_s\ ,\nn
{\cal O}^{(4)}_{13}&=&\cos(\Phi_s)\left[(\partial_x\varphi_s)^2+(\partial_x\bar\varphi_s)^2\right].
\eea

\subsubsection{Terms coupling charge and spin}
Finally, there are eight symmetry-allowed perturbations involving both
spin and charge sectors
\bea
{\cal O}^{(4)}_{15}&=&\cos(\Phi_s)\partial_x\varphi_c\partial_x\bar\varphi_c\ ,\nn
{\cal O}^{(4)}_{16}&=&\cos(\Phi_s)\left[(\partial_x\varphi_c)^2+(\partial_x\bar\varphi_c)^2
\right],\nn
{\cal O}^{(4)}_{17}&=&[(\partial_x\varphi_s)^2+(\partial_x\bar\varphi_s)^2]
[(\partial_x\varphi_c)^2+(\partial_x\bar\varphi_c)^2],\nn
{\cal O}^{(4)}_{18}&=&[(\partial_x\varphi_s)^2+(\partial_x\bar\varphi_s)^2]
\partial_x\varphi_c\partial_x\bar\varphi_c,\nn
{\cal O}^{(4)}_{19}&=&\partial_x\varphi_s\partial_x\bar\varphi_s
[(\partial_x\varphi_c)^2+(\partial_x\bar\varphi_c)^2],\nn
{\cal O}^{(4)}_{20}&=&\partial_x\varphi_s\partial_x\bar\varphi_s\
\partial_x\varphi_c\partial_x\bar\varphi_c,\nn
{\cal O}^{(4)}_{21}&=&[(\partial_x\varphi_s)^2+(\partial_x\bar\varphi_s)^2]\
[\partial_x^2\varphi_c-\partial_x^2\bar\varphi_c],\nn
{\cal O}^{(4)}_{22}&=&\partial_x\varphi_s\partial_x\bar\varphi_s\
[\partial_x^2\varphi_c-\partial_x^2\bar\varphi_c].
\eea

\section{List of irrelevant operators at the LE point}
\label{app:irrLE}
The symmetries of the Hamiltonian at the Luther-Emery point are the same as those listed in Sec. \ref{app:irrops}. Since at the LE point all symmetry-allowed operators in the Hamiltonian  are local in terms of free holons and spinons, we shall give the list of irrelevant operators directly in the fermionic representation. Among the operators listed in Appendix \ref{app:irrops}, those which involve only derivatives of $\varphi_\alpha,\bar \varphi_\alpha$ preserve the same scaling dimension at the LE point; their expressions in the fermionic basis can be obtained straightforwardly by using bosonization identities for spinless fermions. 

On the other hand, operators that contain $\cos(\Phi_s)$ require a more careful analysis. First we note that, taking into account the Klein factors, the  operator is actually  represented by \be
\cos(\Phi_s)\to \eta_{\uparrow}\bar \eta_{\downarrow}\eta_{\downarrow}\bar\eta_{\uparrow}\cos (\Phi_s).\label{fourmajoranas}
\ee
Recall that spin flip and parity act nontrivially on $\eta_\sigma,\bar\eta_\sigma$. However, the product of four Majorana fermions in Eq. (\ref{fourmajoranas}) is invariant under both transformations. As a result, the Klein factors can be safely omitted in the symmetry  analysis of the bosonized perturbations to the Luttinger model. At the LE point, however, $\cos(\Phi_s)$ must be refermionized into free spinons according to Eq. (\ref{spinlessRalpha}). In terms of spinon operators,  spin-flip symmetry is equivalent  to a particle-hole symmetry \be
R^{\phantom\dagger}_s\leftrightarrow R_s^\dagger, \quad L^{\phantom\dagger}_s\leftrightarrow L_s^\dagger,\label{conjugatespinon} \ee
while leaving the spinon Klein factors $\eta_s,\bar \eta_s$ invariant. Parity acts on spinon operators in the form\bea
PR_s(x)P&=&L_s(-x),\quad  PL_s(x)P=R_s(-x),\\
P\eta_sP&=&\bar\eta_s,\quad P\bar\eta_sP= \eta_s.
\eea
The refermionization of $\cos(\Phi_s)$ at the LE point yields\bea
\cos(\Phi_s)&\to& \cos(\Phi_s^*/\sqrt2)\nonumber\\
&\sim &\pi (\eta_s R_s^\dagger \bar \eta_s L^{\phantom\dagger}_s+\eta_s R_s^{\phantom\dagger }\bar \eta_s L^{\dagger}_s)\nonumber\\
&=&-\pi (\eta_s \bar \eta_s R_s^\dagger  L^{\phantom\dagger}_s+ \bar \eta_s \eta_s L^{\dagger}_s R_s^{\phantom\dagger } ).\label{RLtermwithetas}
\eea
Notice that the operator in Eq. (\ref{RLtermwithetas}) is invariant under the spin-flip transformation  (\ref{conjugatespinon}) only if we take the spinon Klein factors into account explicitly. Nevertheless, in the following list of irrelevant operators we shall omit the Klein factors for short and write simply\be
\cos(\Phi_s)\to  (  R_s^\dagger  L^{\phantom\dagger}_s+   L^{\dagger}_s R_s^{\phantom\dagger } ).\label{cosPhiRsLs}
\ee 
Since all the operators that stem from $\cos(\Phi_s)$ contain the same combination of Klein factors in Eq. (\ref{RLtermwithetas}), we can adopt the prescription in Eq. (\ref{cosPhiRsLs}) supplemented by the ad hoc rule that spin-flip symmetry takes $R^{\phantom\dagger}_s\to R^{ \dagger}_s$  but $L^{\phantom\dagger}_s\to -L^{\dagger}_s$.

Importantly, the operator $\cos(\Phi_s)$ has scaling dimension 2 at the SU(2)-symmetric weak coupling regime, but   dimension 1 at the LE point. Therefore, the scaling dimension of  perturbations that contain $\cos(\Phi_s)$ is reduced by 1 as we go from weak coupling to the LE point. For instance, the marginal operator $\mc O^{(2)}_1=\cos(\Phi_s)$ at the SU(2) point becomes the relevant mass term of the Thirring model, $\cos(\Phi_s^*/\sqrt2)\sim R_s^\dagger L^{\phantom\dagger}_s+L_s^\dagger R^{\phantom\dagger}_s$ at the LE point, while the irrelevant (dimension-three) operator $\mc O_7^{(3)}=(\partial_x\bar\varphi_c+\partial_x\varphi_c)\cos(\Phi_s)$ at the SU(2) point gives rise to the marginal spin-charge coupling $(R_s^\dagger L^{\phantom\dagger}_s+L_s^\dagger R^{\phantom\dagger}_s)(R_c^\dagger R^{\phantom\dagger}_c+L_c^\dagger L^{\phantom\dagger}_c)$ at the LE point. This implies that, in order to have the complete list of irrelevant operators up to dimension four at the LE point, we have to consider the refermionization of operators that have dimension 5 at the SU(2) point and were not included in the list in Appendix  \ref{app:irrops}. The latter correspond to operators $\mc V^{(4)}_{21}$ through $\mc V^{(4)}_{27}$ in the list below.

\subsection{Dimension-three operators allowed by symmetry}

\bea
\mc V_1^{(3)}&=&R_c^\dagger\partial_x^2R_c^{\phantom\dagger}+L_c^\dagger\partial_x^2L_c^{\phantom\dagger}+\text{h.c.},\\
\mc V_2^{(3)}&=&R_c^\dagger R_c^{\phantom\dagger}L_c^\dagger i\partial_xL_c^{\phantom\dagger}-L_c^\dagger L_c^{\phantom\dagger} R_c^\dagger i\partial_xR_c^{\phantom\dagger}+\text{h.c.}.\\
\mc V_{3}^{(3)}&=&\partial_x(R_c^\dagger R_c^{\phantom\dagger})L_c^\dagger L_c^{\phantom\dagger}-\partial_x(L_c^\dagger L_c^{\phantom\dagger})R_c^\dagger R_c^{\phantom\dagger},\\
\mc V_{4}^{(3)}&=&\partial_xR_s^\dagger\partial_xL_s^{\phantom\dagger}+\partial_xL_s^\dagger\partial_xR_s^{\phantom\dagger},\\
\mc V_{5}^{(3)}&= &iR_s^\dagger  \partial_xR_s^{\phantom \dagger}L_s^\dagger  R_s^{\phantom\dagger}-iL_s^\dagger  \partial_xL_s^{\phantom \dagger}R_s^\dagger  L_s^{\phantom\dagger}+\text{h.c.},\\
\mc V_6^{(3)}&=&R_c^\dagger R_c^{\phantom\dagger}R_s^\dagger i\partial_xR_s^{\phantom\dagger}-L_c^\dagger L_c^{\phantom\dagger} L_s^\dagger i\partial_xL_s^{\phantom\dagger}+\text{h.c.},\\
\mc V_7^{(3)}&=&R_c^\dagger R_c^{\phantom\dagger}L_s^\dagger i\partial_xL_s^{\phantom\dagger}-L_c^\dagger L_c^{\phantom\dagger} R_s^\dagger i\partial_xR_s^{\phantom\dagger}+\text{h.c.},\\
\mc  V_8^{(3)}&=&(R_c^\dagger R_c^{\phantom\dagger}+L_c^\dagger L_c^{\phantom\dagger} )R_s^\dagger R_s^{\phantom\dagger}L_s^\dagger L_s^{\phantom\dagger},\\
\mc V_{9}^{(3)}&=&(iR_c^\dagger  \partial_xR_c^{\phantom \dagger}-iL_c^\dagger  \partial_xL_c^{\phantom \dagger}+\text{h.c.})\nonumber\\
&&\times(R_s^\dagger   L_s^{\phantom \dagger}+L_s^\dagger R_s^{\phantom \dagger}),\\
\mc  V_{10}^{(3)}&=&R_c^\dagger   R_c^{\phantom \dagger}L_c^\dagger L_c^{\phantom \dagger}(R_s^\dagger   L_s^{\phantom \dagger}+L_s^\dagger R_s^{\phantom \dagger}),\\
\mc V_{11}^{(3)}&=&\partial_x(R_c^\dagger   R_c^{\phantom \dagger}-L_c^\dagger L_c^{\phantom \dagger})(R_s^\dagger   L_s^{\phantom \dagger}+L_s^\dagger  R_s^{\phantom \dagger}).
\eea

\subsection{Dimension-four operators allowed by symmetry}

\subsubsection{Charge sector only}
\bea
\mc V^{(4)}_1&=&iR_c^\dagger\partial_x^3R_c^{\phantom\dagger}-iL_c^\dagger\partial_x^3L_c^{\phantom\dagger}+\text{h.c.},\\
\mc V^{(4)}_2&=&R_c^\dagger \partial_xR_c^{\phantom\dagger}L_c^\dagger \partial_xL_c^{\phantom\dagger}+\text{h.c.},\\
\mc V^{(4)}_3&=&R_c^\dagger \partial^2_xR_c^{\phantom\dagger}L_c^\dagger L_c^{\phantom\dagger}+L_c^\dagger \partial^2_xL_c^{\phantom\dagger}R_c^\dagger R_c^{\phantom\dagger}+\text{h.c.},\\
\mc V^{(4)}_4&=&R_c^\dagger \partial_xR_c^\dagger R_c^{\phantom\dagger} \partial_x R_c^{\phantom\dagger}+L_c^\dagger \partial_xL_c^\dagger L_c^{\phantom\dagger} \partial_x L_c^{\phantom\dagger},\\
\mc V^{(4)}_5&=& \partial_x(R_c^\dagger  R_c^{\phantom\dagger}) \partial_x( L_c^\dagger  L_c^{\phantom\dagger}),\\
\mc V^{(4)}_{6}&=&\partial_x(R_c^\dagger R_c^{\phantom\dagger})L_c^\dagger i\partial_xL_c^{\phantom\dagger}+\partial_x(L_c^\dagger L_c^{\phantom\dagger})R_c^\dagger i\partial_xR_c^{\phantom\dagger}\nonumber\\
&&+\text{h.c.}.
\eea

\subsubsection{Spin sector only}
\bea
\mc V^{(4)}_{7}&=&iR_s^\dagger\partial_x^3R_s^{\phantom\dagger}-iL_s^\dagger\partial_x^3L_s^{\phantom\dagger},\\
\mc V^{(4)}_{8}&=&R_s^\dagger\partial_xR_s^{\phantom\dagger}L_s^\dagger\partial_xL_s^{\phantom\dagger}+\text{h.c.},\\
\mc V^{(4)}_{9}&=&R_s^\dagger\partial_xR_s^\dagger L_s^{\phantom\dagger}\partial_xL_s^{\phantom\dagger}+L_s^\dagger\partial_xL_s^\dagger R_s^{\phantom\dagger}\partial_xR_s^{\phantom\dagger},\\
\mc V^{(4)}_{10}&=&R_s^\dagger\partial_x^2R_s^{\phantom\dagger}L_s^\dagger L_s^{\phantom\dagger}+L_s^\dagger\partial_x^2L_s^{\phantom\dagger}R_s^\dagger R_s^{\phantom\dagger}+\text{h.c.},\\
\mc V^{(4)}_{11}&= &R_s^\dagger  \partial_xR_s^\dagger R_s^{\phantom \dagger} \partial_xR_s^{\phantom \dagger}+L_s^\dagger  \partial_xL_s^\dagger L_s^{\phantom \dagger} \partial_xL_s^{\phantom \dagger},\\
\mc V^{(4)}_{12}&= &\partial_x(R_s^\dagger R_s^{\phantom\dagger})\partial_x(L_s^\dagger L_s^{\phantom\dagger}).
\eea

\subsubsection{Terms coupling charge and spin}
\bea
\mc V^{(4)}_{13}&=&(iR_c^\dagger \partial_x R_c^{\phantom \dagger}+\text{h.c.})(iR_s^\dagger \partial_x R_s^{\phantom \dagger}+\text{h.c.})\nonumber\\
&&+(iL_c^\dagger \partial_x L_c^{\phantom \dagger}+\text{h.c.})(iL_s^\dagger \partial_x L_s^{\phantom \dagger}+\text{h.c.}),\\
\mc V^{(4)}_{14}&=&(iR_c^\dagger \partial_x R_c^{\phantom \dagger}+\text{h.c.})(iL_s^\dagger \partial_x L_s^{\phantom \dagger}+\text{h.c.})\nonumber\\
&&+(iL_c^\dagger \partial_x L_c^{\phantom \dagger}+\text{h.c.})(iR_s^\dagger \partial_x R_s^{\phantom \dagger}+\text{h.c.}),\\
\mc V^{(4)}_{15}&=&R_c^\dagger   R_c^{\phantom \dagger}L_c^\dagger L_c^{\phantom \dagger}(iR_s^\dagger \partial_x R_s^{\phantom \dagger}-iL_s^\dagger \partial_x L_s^{\phantom \dagger}+\text{h.c.}),\\
\mc V^{(4)}_{16}&=&(iR_c^\dagger \partial_x R_c^{\phantom \dagger}-iL_c^\dagger \partial_x L_c^{\phantom \dagger}+\text{h.c.})R_s^\dagger   R_s^{\phantom \dagger}L_s^\dagger   L_s^{\phantom \dagger},\\
\mc V^{(4)}_{17}&=&R_c^\dagger   R_c^{\phantom \dagger}L_c^\dagger L_c^{\phantom \dagger}R_s^\dagger   R_s^{\phantom \dagger}L_s^\dagger L_s^{\phantom \dagger},\\
\mc V^{(4)}_{18}&=&\partial_x(R_c^\dagger   R_c^{\phantom \dagger})(iR_s^\dagger \partial_x R_s^{\phantom \dagger}+\text{h.c.})\nonumber\\
&&+\partial_x(L_c^\dagger L_c^{\phantom \dagger})(iL_s^\dagger \partial_x L_s^{\phantom \dagger}+\text{h.c.}),\\
\mc V^{(4)}_{19}&=&\partial_x(R_c^\dagger   R_c^{\phantom \dagger})(iL_s^\dagger \partial_x L_s^{\phantom \dagger}+\text{h.c.})\nonumber\\
&&+\partial_x(L_c^\dagger L_c^{\phantom \dagger})(iR_s^\dagger \partial_x R_s^{\phantom \dagger}+\text{h.c.}),\\
\mc V^{(4)}_{20}&=&\partial_x(R_c^\dagger   R_c^{\phantom \dagger}-L_c^\dagger L_c^{\phantom \dagger})R_s^\dagger   R_s^{\phantom \dagger}L_s^\dagger  L_s^{\phantom \dagger},
\eea
\bea
\mc V^{(4)}_{21}&=&(R_c^\dagger\partial_x^2R_c^{\phantom\dagger}+L_c^\dagger\partial_x^2L_c^{\phantom\dagger}+\text{h.c.})\nonumber\\
&&\times(R_s^\dagger   L_s^{\phantom \dagger}+L_s^\dagger  R_s^{\phantom \dagger}),\\
\mc V^{(4)}_{22}&=&(R_c^\dagger R_c^{\phantom\dagger}L_c^\dagger i\partial_xL_c^{\phantom\dagger}-L_c^\dagger L_c^{\phantom\dagger} R_c^\dagger i\partial_xR_c^{\phantom\dagger}+\text{h.c.}) \nonumber\\
&&\times(R_s^\dagger   L_s^{\phantom \dagger}+L_s^\dagger  R_s^{\phantom \dagger})\\
\mc V^{(4)}_{23}&=&R_c^\dagger R_c^{\phantom\dagger}R_s^\dagger i\partial_xR_s^{\phantom\dagger}L_s^\dagger  R_s^{\phantom \dagger}\nonumber\\
&&-L_c^\dagger L_c^{\phantom\dagger} L_s^\dagger i\partial_xL_s^{\phantom\dagger}R_s^\dagger   L_s^{\phantom \dagger}+\text{h.c.},\\
\mc V^{(4)}_{24}&=&R_c^\dagger R_c^{\phantom\dagger}L_s^\dagger i\partial_xL_s^{\phantom\dagger}R_s^\dagger   L_s^{\phantom \dagger}\nonumber\\
&&-L_c^\dagger L_c^{\phantom\dagger} R_s^\dagger i\partial_xR_s^{\phantom\dagger}L_s^\dagger  R_s^{\phantom \dagger}+\text{h.c.}\\
\mc V^{(4)}_{25}&=&\partial_x^2(R^\dagger_cR_c^{\phantom\dagger}+L^\dagger_cL_c^{\phantom\dagger})(R^\dagger_sL_s^{\phantom\dagger}+L_s^\dagger R_s^{\phantom\dagger}),\\
\mc V^{(4)}_{26}&=&\partial_x(R^\dagger_ci\partial_xR_c^{\phantom\dagger}+L^\dagger_ci\partial_xL_c^{\phantom\dagger})(R^\dagger_sL_s^{\phantom\dagger}+L_s^\dagger R_s^{\phantom\dagger}),\\
\mc V^{(4)}_{27}&=&[\partial_x(R_c^\dagger R_c^{\phantom\dagger})L_c^\dagger L_c^{\phantom\dagger}-\partial_x(L_c^\dagger L_c^{\phantom\dagger})R_c^\dagger R_c^{\phantom\dagger}]\nonumber\\
&&\times(R^\dagger_sL_s^{\phantom\dagger}+L_s^\dagger R_s^{\phantom\dagger}).
\eea


\section{Constraining the mobile impurity model in the SU(2)-symmetric case\label{app:Su2}}

We can  impose SU(2) symmetry in the parameters of the mobile impurity model of Section \ref{sec:MIMaway} by requiring that the edge exponents of longitudinal and transverse spin-spin correlations coincide\cite{work7}.  First consider the longitudinal component of the spin density operator:\bea
S^z(x)&\sim &R_{\uparrow}^\dagger(x)R_{\uparrow}^{\phantom\dagger}(x)-R_{\downarrow}^\dagger(x)R_{\downarrow}^{\phantom\dagger}(x)\nonumber\\
&\sim &\mc O_s^\dagger(x)\mc O_s^{\phantom\dagger}(x),
\eea
where the charge strings are cancelled in the sense of the lowest order in the OPE. We then project  $\mc O_s(x)$ so as to  create a high-energy spinon:\be
\mc O_s(x)\sim e^{-iqx}\chi_s^\dagger(x)e^{\frac{i}{4\sqrt2}\Phi_s^*(x)},
\ee
while $\mc O_s^\dagger(x)$ acts only in the low-energy subband:\be
\mc O_s^\dagger(x)\sim e^{\frac{i}{\sqrt2}\varphi_s^*-\frac{i}{4\sqrt2}\Phi_s^*(x)}.
\ee
Cancelling the neutral string, we obtain\bea
S^z(x)&\sim& e^{-iqx} \chi_s^\dagger (x)e^{\frac{i}{\sqrt2}\varphi_s^*(x)}.
\eea
In terms of the transformed impurity field\be
S^z(x)\sim e^{-iqx} d_s^\dagger(x)e^{ i\left(\frac{1}{\sqrt2}+\gamma_s\right)\varphi_s^*+i\bar \gamma_s\bar\varphi_s^*+i  \gamma_c \varphi_c^*+i\bar \gamma_c\bar\varphi_c^*}.\label{repSz}
\ee

Now, consider  the transverse component:\bea
S^-(x)&\sim &R^\dagger_{\downarrow}(x)R^{\phantom\dagger}_{\uparrow}(x)\nonumber\\
&\sim&\mc O_s(x)\mc O_s(x)\nonumber\\
&\sim &e^{-iqx}\chi_s^\dagger(x)e^{-\frac{i}{2\sqrt2}\varphi_s^*(x)+\frac{i}{2\sqrt2}\bar\varphi_s^*(x)}\nonumber\\
&=&e^{-iqx}d_s^\dagger(x)e^{-i\left(\frac{1}{2\sqrt2}-\gamma_s\right)\varphi_s^*+i\left(\frac{1}{2\sqrt2}+\bar\gamma_s\right)\varphi_s^*}\nonumber\\
&&\ \times e^{i  \gamma_c \varphi_c^*+i\bar \gamma_c\bar\varphi_c^*}.\label{repSminus}
\eea
Given the expressions \fr{repSz} and \fr{repSminus} we can calculate
threshold exponents in the Fourier transforms of the spin correlations
functions $\langle S^z(x,t)S^z(0,0)\rangle$ and $\langle
S^+(x,t)S^-(0,0)\rangle$ and impose that they share the same exponents
in the SU(2)-symmetric case. This has to be the case even for
higher harmonics, taking into account backscattering
processes\cite{work7}.  A shortcut is to compare the scaling
dimensions of the strings in Eqs. (\ref{repSz}) 
and (\ref{repSminus}). We must have\bea 
\frac{1}{\sqrt2}+\gamma_s&=&\frac{1}{2\sqrt2}-\gamma_s,\\
-\bar \gamma_s&=&\frac{1}{2\sqrt2}+\bar \gamma_s.
\eea
It follows that  SU(2) symmetry imposes \be
\gamma_s=\bar\gamma_s=-\frac1{4\sqrt2}. \label{gammasSU2}
\ee
These values of $\gamma_s,\bar \gamma_s$ are of order 1, as expected since the spinons are strongly interacting at the SU(2) point. Moreover,   these values  are consistent with the exact result for the Hubbard model at zero magnetic field (see Sec. \ref{sec:Hubbard}). Using Eq. (\ref{gammasSU2}) and setting $K_s=1$ in Eq. (\ref{nus}) yields \be
\nu_{s,+}=  \nu_{s,-}=0. \ee

\section{Friedel oscillations} 
\label{app:parity}
In this appendix we summarize some useful facts regarding Friedel
oscillations for open boundary conditions.
At low energies the charge density operator has an expansion of the form
\be
n_j\sim \sum_{n=0}\rho_{2nk_F}(x),
\ee
where $\rho_{2nk_F}(x)$ denotes the Fourier components with
momenta $\pm 2nk_F$. For periodic boundary conditions it follows
from \fr{opexpansion} and the site-parity symmetry \fr{siteparity} that
the leading contributions to the $2k_F$ and $4k_F$ components are
\bea
\rho_{2k_F}(x)&=& \widehat{A}_{2k_F}
\sin\left(\frac{\Phi_c}{2}-2k_Fx\right)\ \cos\left(\frac{\Phi_s}{2}\right)
+\ldots,\nn
\rho_{4k_F}(x)&=& \widehat{A}_{4k_F}
\cos\left(\Phi_c-4k_Fx\right)+\ldots,
\label{rhos}
\eea
where $\widehat{A}_{2k_F}$ and $\widehat{A}_{4k_F}$ are non-universal
amplitudes that include Klein factors. The leading contributions to
zero momentum component are
\bea
\rho_{0}(x)&=&n-\frac{1}{2\pi}\partial_x\Phi_c +
\widehat{A}_{0}\partial_x\Phi_c\ \cos\left(\frac{\Phi_s}{2}\right)
+\ldots,\nn
\eea
where $n$ is the band filling.

\subsection{Open boundary conditions}
For open boundary conditions we still have expansions like
\fr{opexpansion}, but the chiral Bose fields no longer commute and hence the
Klein factors $\Gamma_{n,m}^\sigma$ need to be adjusted accordingly.
The mode expansions for the chiral Bose fields are 
\bea
\varphi_c(x)&=&\frac{a+\pi_0}{2}
+\frac{x\varphi_0}{2(L+1)}\nn
&&+i\sum_{n=1}^\infty \sqrt{\frac{2}{n}}\left[e^{-iq_n x}\alpha_n-{\rm h.c.}\right],\nn
\bar\varphi_c(x)&=&\frac{a-\pi_0}{2}+\frac{x\varphi_0}{2(L+1)}\nn
&&-
i\sum_{n=1}^\infty \sqrt{\frac{2}{n}}\left[e^{iq_n x}\alpha_n-{\rm h.c.}\right],
\label{modes}
\eea
where $\lbrack\pi_0,\varphi_0\rbrack=8\pi i$ and $q_n=\pi
n/(L+1)$. The commutator between different chiralities is
\be
[\varphi_\alpha(x),\bar\varphi_\alpha(y)]=
\begin{cases}
0 & \text{if } x=y=0,\\
2\pi i & \text{if\ } 0<x,y<L+1,\\
4\pi i & \text{if } x=y=L+1.
\end{cases}
\ee
The boundary conditions on the Fermi creation and
annihilation operators are
\be
c_{j=0,\sigma}=0=c_{j=L+1,\sigma}.
\label{bcs}
\ee
Imposing the boundary conditions \fr{bcs} on the bosonized expression
\fr{opexpansion} (with Klein factors appropriate for the mode expansions \fr{modes})
leads to conditions of the form
\bea
\Phi_c(0)|\psi_0\rangle&=&c_1|\psi_0\rangle\ ,\quad\Phi_c(L+1)|\psi_0\rangle=c_2|\psi_0\rangle\ ,\nn
\Phi_s(0)|\psi_0\rangle&=&\Phi_s(L+1)|\psi_0\rangle=c'_1|\psi_0\rangle\ .
\eea
Importantly, the actual values of the $c_{1,2}$ and $c_{1,2}'$ depend on all
the amplitudes $A_{n,m}$ in \fr{opexpansion} and are thus non-universal.

\subsection{Friedel Oscillations}
For open boundary conditions the $2k_F$ and $4k_F$ components of the
total charge density are again given by expressions of the form \fr{rhos}.
Using the mode expansions we then can determine the form of the
Friedel oscillations
\bea
\langle\psi_0|\rho_{2k_F}(x)|\psi_0\rangle
&\sim&B_{2k_F}\frac{\sin\Big(2k_F'x-\frac{c_1}{2}\Big)}
{\Big(\frac{2}{\epsilon}\sin\big(\frac{\pi
    x}{L+1}\big)\Big)^{(1+K_c)/2}}\ ,\nn
\langle\psi_0|\rho_{4k_F}(x)|\psi_0\rangle&\sim&
B_{4k_F}\frac{\cos\big(4k_F'x-c_1\big)}
{\Big(\frac{2}{\epsilon}\sin\big(\frac{\pi
    x}{L+1}\big)\Big)^{2K_c}},
\eea
where $\epsilon$ is a short-distance cutoff, we have taken into
account $K_c\neq 1$ and defined
\be
k_F'=k_F+\frac{c_1-c_2}{4(L+1)}\ .
\ee
The upshot is that Friedel oscillations involve two \emph{non-universal}
interaction-dependent parameters: the phase shift $c_1$ and the $1/L$
shift of $k_F$.


\begin{thebibliography}{12}

\bibitem{xu}C. Xu and S. Sachdev, Phys. Rev. Lett. {\bf105}, 057201 (2010). 
\bibitem{deshpande}V. V. Deshpande, M.  Bockrath, L. I.  Glazman, and A. Yacoby,  Nature {\bf464}, 209  (2010).
\bibitem{auslaender}O. M. Auslaender {\it et al.}, Science {\bf295}, 825 (2002).
\bibitem{kim}B. J. Kim {\it et al.}, Nat. Phys. {\bf2}, 397  (2006). 
\bibitem{jompol}Y. Jompol {\it et al}., Science {\bf31}, 597 (2009).
\bibitem{schlappa}J. Schlappa {\it et al.},  Nature {\bf485}, 82 (2012). 
\bibitem{hubbardbook}F. H. L. Essler, H. Frahm, F. G\"ohmann,
  A. Kl\"umper, and V. E. Korepin, {\it The One Dimensional Hubbard Model} (Cambridge University Press, Cambridge, 2005).

\bibitem{EK94a}
F. H. L. Essler and V. E. Korepin, Phys. Rev. Lett. {\bf72}, 908 (1994).

\bibitem{EK94b}
F. H. L. Essler and V. E. Korepin, Nucl. Phys. {\bf B426}, 505 (1994).

\bibitem{Andrei}
N. Andrei, arXiv:cond-mat/9408101.

\bibitem{Deguchi}
T. Deguchi, F. H. L. Essler, F. G\"ohmann, V.E. Korepin, A. Kl\"umper, 
and K. Kusakabe, Phys. Rep. {\bf 331}, 197 (2000).

\bibitem{penc1}
K. Penc, F. Mila, and H. Shiba, Phys. Rev. Lett. {\bf 75}, 894-897 (1995).

\bibitem{favand}
J. Favand, S. Haas, K. Penc, F. Mila, and E. Dagotto, Phys. Rev. B {\bf
  55}, R4859 (1997);

\bibitem{penc}K. Penc, K. Hallberg, F. Mila, and H. Shiba,
  Phys. Rev. B {\bf55}, 15475 (1997).

\bibitem{benthien}
H. Benthien, F. Gebhard, and E. Jeckelmann, Phys. Rev. Lett.
{\bf 92}, 256401 (2004).

\bibitem{adrian}
A.E. Feiguin and D.A. Huse, Phys. Rev. B{\bf 79}, 100507(R) (2009).

\bibitem{arikawa01}
M. Arikawa, Y. Saiga, and Y. Kuramoto, Phys. Rev. Lett. {\bf 86}, 3096 (2001).

\bibitem{arikawa04}
M. Arikawa, T. Yamamoto, Y. Saiga, and Y. Kuramoto, Nucl. Phys. {\bf
  B702}, 380 (2004).

\bibitem{Zam}
A.~B. Zamolodchikov and Al.~B. Zamolodchikov, Ann. Phys. {\bf 120},  253 (1979).

\bibitem{Fadd}
L.~D. Faddeev, Sov. Sci. Rev. Math. Phys.~C {\bf 1},  107  (1980).


\bibitem{kareljan}
R. A. J. van Elburg and K. Schoutens, Phys. Rev. B{\bf 58}, 15704 (1998).

\bibitem{JNW}
G. I. Japaridze, A. A. Nersesyan, and P. B. Wiegmann, Nucl. Phys. {\bf
  B230}, 511 (1984).

\bibitem{woynarovich}
F. Woynarovich, J. Phys. \textbf{A 22}, 4243 (1989).

\bibitem{frahmkorepin}
H.~Frahm and V.~E. Korepin, Phys. Rev. B \textbf{42}, 10553 (1990).

\bibitem{Affleck}
I. Affleck, in {\em Fields, Strings and Critical Phenomena},
eds E. Br\'ezin and J. Zinn-Justin, (Elsevier, Amsterdam, 1989).

\bibitem{gogolin}A. O. Gogolin, A. A. Nersesyan, and A. M. Tsvelik,
  {\it Bosonization and Strongly Correlated Systems} (Cambridge
  University Press, 
Cambridge, 1999).

\bibitem{giamarchi}T. Giamarchi, {\it Quantum Physics in One
  Dimension} (Claredon Press, Oxford, 2004).

\bibitem{work1}
M. Pustilnik, M. Khodas, A. Kamenev, and L.I. Glazman,
Phys. Rev. Lett. {\bf 96}, 196405 (2006).
\bibitem{rozhkov}A. Rozhkov, Eur. Phys. J. B {\bf47}, 193 (2005).

\bibitem{Roz06}
A. V. Rozhkov, Phys. Rev. B~{\bf 74}, 245123 (2006).

\bibitem{carmelo1}
J. M. P. Carmelo, K. Penc, L.M. Martelo, P. D. Sacramento, J. M. B. Lopes
Dos Santos, R. Claessen, M. Sing and U. Schwingenschl\"ogl,
Europhys. Lett. {\bf 67}, 233 (2004).

\bibitem{carmelo2}
J. M. P. Carmelo, K. Penc, P. D. Sacramento, M. Sing and R. Claessen,
J. Phys.: Condens. Matter {\bf 18}, 5191 (2006).

\bibitem{carmelo3}
J. M. P. Carmelo, D. Bozi and K. Penc, J. Phys. Cond. Mat. {\bf 20},
415103 (2008).


\bibitem{work2}
R. G. Pereira, J. Sirker, J.-S. Caux, R. Hagemans, J. M. Maillet,
S. R. White, and I. Affleck, Phys. Rev. Lett. {\bf 96}, 257202 (2006).
\bibitem{BAW1}
E. Bettelheim, A.~G. Abanov, and P. Wiegmann,
Phys. Rev. Lett. {\bf 97}, 246401 (2006).
\bibitem{work3}
M. Khodas, M. Pustilnik, A. Kamenev, and L.I. Glazman,
Phys. Rev. Lett. {\bf 99}, 110405 (2007).

\bibitem{work4}
M. Khodas, M. Pustilnik, A. Kamenev, and L.I. Glazman,
Phys. Rev. B~{\bf 76}, 155402 (2007).
\bibitem{work5}
R. G. Pereira, S. R. White, and I. Affleck, 
Phys. Rev. Lett. {\bf 100}, 027206 (2008).
\bibitem{IG08}
A. Imambekov and L.I. Glazman, Phys. Rev. Lett. {\bf 100}, 206805
(2008).
\bibitem{work6}
V. V.  Cheianov and M. Pustilnik, Phys. Rev. Lett. {\bf 100}, 126403
(2008).

\bibitem{BAW2}
E. Bettelheim, A. G. Abanov, and P. Wiegmann, J. Phys. {\bf A41},
392003 (2008). 

\bibitem{ABW}
A.~G. Abanov, E. Bettelheim, and P. Wiegmann, J. Phys. {\bf A42}, 135201
(2009). 

\bibitem{work7}
A. Imambekov and L. I. Glazman, Phys.Rev. Lett.~{\bf 102}, 126405
(2009).

\bibitem{work8}
A. Imambekov and L. I. Glazman, Science {\bf 323}, 228 (2009).

\bibitem{PWA09}
R. G. Pereira, S. R. White, and I. Affleck, 
Phys. Rev. B~{\bf 79}, 165113 (2009).

\bibitem{pereira12}
R. G. Pereira, Int. J. Mod. Phys. {\bf B~26}, 1244008 (2012).

\bibitem{Roz14}
A. V. Rozhkov, Phys. Rev. Lett. {\bf 112}, 106403 (2014).

\bibitem{austen}
T. Price and A. Lamacraft, Phys. Rev. B{\bf 90}, 241415 (2014). 

\bibitem{PS}R. G. Pereira and E. Sela, Phys. Rev. B 82, 115324 (2010).

\bibitem{Pereira}
R. G. Pereira, K. Penc, S. R. White, P. D. Sacramento, J.~M.~P. Carmelo,
Phys. Rev. B~{\bf 85}, 165132 (2012).

\bibitem{SIG1}
T. L. Schmidt, A. Imambekov, and L. I. Glazman, Phys. Rev. Lett. {\bf
  104}, 116403 (2010).
\bibitem{SIG2}
T. L. Schmidt, A. Imambekov, and L. I. Glazman,  Phys. Rev. B~{\bf 82}
245104 (2010).


\bibitem{FHLE}
F.~H.~L.~Essler, Phys. Rev. B~{\bf 81}, 205120 (2010).

\bibitem{SIG3}
A. Imambekov, T. L. Schmidt, and L. I. Glazman,  Rev. Mod. Phys {\bf 84},
1253 (2012).

\bibitem{ts}
O. Tsyplyatyev and A. J. Schofield, Phys. Rev. B~{\bf 90}, 014309
(2014).

\bibitem{Seabra}
L. Seabra, F. H. L. Essler, F. Pollmann, I. Schneider, and T. Veness,
Phys. Rev. B {\bf 90}, 245127 (2014).

\bibitem{korepinbook}V. E. Korepin, N. M. Bogoliubov, and A. G. Izergin, {\it Quantum Inverse Scattering Method and Correlation Functions} (Cambridge University Press, Cambridge, 1999).

\bibitem{Faddeev84}
L. D. Faddeev and L. Takhtajan, Jour. Sov. Math. {\bf 24}, 241
(1984).

\bibitem{review} F. H. L. Essler and R. M. Konik, in \emph{From Fields to Strings: Circumnavigating Theoretical Physics}, ed. M. Shifman, A. Vainshtein and J. Wheater (World Scientific, Singapore, 2005) and
references therein; cond-mat/0412421.

\bibitem{Schotte}
K. D. Schotte and U. Schotte, Phys. Rev. {\bf 182}, 479 (1969).


\bibitem{luther}A. Luther and V. J. Emery, Phys. Rev. Lett. {\bf33}, 589 (1974).


\bibitem{haldane}
F. D. M. Haldane, Phys. Rev. Lett. {\bf47}, 1840 (1982).

\bibitem{frahmkorepin2}
H.~Frahm and V.~E. Korepin, Phys. Rev. B {\bf 43}, 5653 (1991).

\bibitem{AEM}
D. B. Abraham, F. H. L. Essler, and A. Maciolek, Phys. Rev. Lett. {\bf
  98}, 170602 (2007). 

\bibitem{coleman}
S. Coleman, Phys. Rev. D{\bf 11}, 2088, (1975).

\bibitem{ZCG}
M. B. Zvonarev, V. V. Cheianov, and T. Giamarchi, J. Stat. Mech. P07035
(2009).

\bibitem{karimi}
H. Karimi and I. Affleck, Phys. Rev. B{\bf 84}, 174420 (2011).

\bibitem{ludwig}
C.G. Callan, I.R. Klebanov, A.W.W. Ludwig and  J.M. Maldacena,
Nucl.Phys. {\bf B422}, 417 (1994).


\bibitem{xiang}
T. Xiang, Phys. Rev. B {\bf53}, R10445(R) (1996).

\bibitem{schollwock}
U. Schollw\"ock, Rev. Mod. Phys. {\bf77}, 259  (2005).


\bibitem{EKS1}
F. H. L. Essler, V. E. Korepin, and K. Schoutens, Phys. Rev. Lett. {\bf 67},
3848 (1991).

\bibitem{EKS2}
F. H. L. Essler, V. E. Korepin, and K. Schoutens, Nucl. Phys. {\bf B372},
559 (1992).


\bibitem{ogatashiba}M. Ogata and H. Shiba, Phys. Rev. B \textbf{41}, 2326 (1990).






\bibitem{Ludwig2003} A.~W.~W.~Ludwig and K.~J.~Wiese, Nucl. Phys. B \textbf{661}, 577 (2003).

\bibitem{Eggert1996} S.~Eggert, Phys.~Rev.~B \textbf{54}, R9612 (1996). 

\bibitem{Nomura}
K. Okamoto and K. Nomura, Phys. Lett. {\bf A169}, 433 (1992).


\bibitem{soeffing2009} S.A. S\"offing, M. Bortz, I. Schneider, A. Struck, M. Fleischhauer, and S. Eggert, Phys.~Rev.~B \textbf{79}, 195114 (2009).

\bibitem{DMRG} 
S.~R. White, Phys. Rev. Lett. \textbf{69}, 2863 (1992).

\bibitem{DMRG2} 
S.~R. White, Phys. Rev. B \textbf{48}, 10345 (1993).

\end{thebibliography}
\end{document}